\documentclass[%
 aip,
 amsmath,amssymb,
reprint,%
]{revtex4-1}

\usepackage{graphicx}
\usepackage{dcolumn}
\usepackage{bm}
\usepackage{xcolor}

\usepackage[utf8]{inputenc}
\usepackage[T1]{fontenc}
\usepackage{mathptmx}
\usepackage{float}

\begin{document}

\preprint{AIP/123-QED}

\title[]{\textcolor{black}{Effect of rate of change of parameter on early warning signals for critical transitions}}

\author{Induja Pavithran}
 \affiliation{Department of Physics, IIT Madras, Chennai-600036, India}
\author{R. I. Sujith}%
\email{sujith@iitm.ac.in}
\affiliation{ Department of Aerospace Engineering, IIT Madras, Chennai-600036, India}

\date{\today}
             
\begin{abstract}
Many dynamical systems exhibit abrupt transitions or tipping as the control parameter is varied. In scenarios where the parameter is varied continuously, the rate of change of control parameter \textcolor{black}{greatly affects the performance of early warning signals (EWS) for such critical transitions.}
We study the impact of variation of the control parameter with a finite rate on the performance of \textcolor{black}{EWS for critical transitions} in a thermoacoustic system (a horizontal Rijke tube) exhibiting subcritical Hopf bifurcation. There is a growing interest in developing early warning signals for tipping in real systems. Firstly, we explore the efficacy of early warning signals based on critical slowing down and fractal \textcolor{black}{characteristics}. From this study, lag-1 autocorrelation (AC) and Hurst exponent ($H$) are found to be good measures to \textcolor{black}{predict the transition} well-before the tipping point. 
The warning time, obtained using AC and $H$, reduces with an increase in the rate of change of the control parameter following an inverse power law relation. Hence, for very fast rates, the warning time may be too short to perform any control action. Furthermore, we report the observation of a hyperexponential scaling relation between the AC and the variance of fluctuations during such dynamic Hopf bifurcation. We construct a theoretical model for noisy Hopf bifurcation wherein the control parameter is continuously varied \textcolor{black}{at different rates to study the effect of rate of change of parameter on EWS}. Similar results, including the hyperexponential scaling, are observed in the model as well.

\end{abstract}

\maketitle

\begin{quotation}

Critical transitions, which can result in sudden devastating changes to the state of the system, are ubiquitous in natural, economic, and social systems. \textcolor{black}{Continuous variation of control parameter in time can delay the transition due to memory effects. Therefore,} the rate of change of control parameter as well as the value of the control parameter determine the tipping point. In such cases, \textcolor{black}{we observe a rate-dependent tipping-delay from the critical point predicted by bifurcation analysis}. Furthermore, the interplay between the inherent noise in the system and the rate of change of parameter blurs the stability margins. Therefore, it is essential to develop effective early warnings signals (EWS) to predict \textcolor{black}{such transitions in practical systems}. In recent years, many studies have explored various EWS to predict the onset of critical transitions. We conduct experiments in a thermoacoustic system exhibiting \textcolor{black}{Hopf bifurcation. In a  thermoacoustic system, the positive feedback between the unsteady heat release rate fluctuations and the acoustic field in the confinement can result in a transition to high amplitude limit cycle oscillations. These self-sustained large amplitude oscillations are dangerous to the system}. For different rates of change of the control parameter, we compare the efficacy of various early warning signals \textcolor{black}{based on critical slowing down and fractal characteristics of the signals acquired from a thermoacoustic system}. We also investigate the variation of warning time with the rate of change of the parameter.

\end{quotation}

\section{\label{sec1}Introduction}

The dynamics of many natural and human-made systems are controlled by various parameters which evolve in time. A tiny perturbation in a system parameter can qualitatively alter the state of the system when it crosses a critical threshold. This phenomenon is generally known as tipping or critical transition, wherein a small change of a parameter can cause a sudden transition in the state of the system \cite{lenton2008tipping}. The term tipping has been used to explain various phenomena such as phase transitions and bifurcations, which are often associated with dangerous and catastrophic transitions. Tipping is observed in real systems such as climate systems \cite{lenton2008tipping}, ecological systems \cite{scheffer2001catastrophic}, financial markets \cite{sornette1997large} and biological systems \cite{venegas2005self, litt2001epileptic, mcsharry2003prediction}.
In earth's climate system, a gradual change in local climate can affect ecosystems and can sometimes trigger a drastic switch to a contrasting state \cite{scheffer2001catastrophic,carpenter1999management}. 
Contagion in financial markets \cite{national2007new, may2008ecology}, spontaneous asthma attacks \cite{venegas2005self}, and epileptic seizures \cite{litt2001epileptic, mcsharry2003prediction} are other instances of tipping.

There are various mechanisms through which tipping occurs. Recently, Ashwin $et\ al.$ \cite{ashwin2012tipping} classified the underlying mechanisms as bifurcation-induced (B), noise-induced (N) or rate-induced (R) tipping. B-tipping occurs when a system parameter is varied slowly through the bifurcation point, commonly known as the slow passage through the bifurcation. Examples of B-tipping include saddle-node, transcritical, pitchfork and Hopf bifurcations. In N-tipping, the system can jump to another stable state due to the presence of noise of sufficient amplitude. In other words, noise can drive the system between the coexisting attractors in systems exhibiting multistability.

Sometimes, the rate of change of parameter plays a more pivotal role than the actual value of the parameter. Ashwin et al. \cite{ashwin2012tipping} showed that when a \textcolor{black}{rate-sensitive }parameter is varied as a function of time, at a slow rate, the system dynamics follows the \textcolor{black}{quasi-static attractor}. For faster rates of change of the parameter, above a critical rate, they observed that the system can be driven outside the basin of attraction of the quasi-static attractor, and can evolve towards a new stable state resulting in rate induced tipping (R-tipping). On the other hand, by varying the bifurcation parameter in a bistable system, one can achieve preconditioned rate induced tipping, as demonstrated by Tony et al. \cite{tony2017experimental}. They reported that the system could be driven towards the basin of attraction of limit cycle before the actual loss of stability of fixed point, for fast enough rates with a finite amplitude initial perturbation. Here, the tipping depends on the rate of change of control parameter and initial conditions. \textcolor{black}{In these cases, the rate at which the parameter is varied determines the tipping point, not the absolute value of parameter. However, practical systems may exhibit tipping phenomena as a result of a combination of bifurcation, rate and noise}. 

\textcolor{black}{In the current study, we explore the effects of rates of change of bifurcation parameter on `bifurcation induced tipping'. When we vary the bifurcation parameter continuously at a finite rate,} tipping will be delayed considerably from the parameter value predicted by the bifurcation analysis \cite{baer1989slow}. Due to the continuous variation of the control parameter, the system stays in the vicinity of the unstable attractor for some time even after the stability is lost. \textcolor{black}{This phenomenon is referred to as `rate-delayed tipping' \cite{bonciolini2018experiments}}. The delay observed in the transition is found to be dependent on the rate of change of parameter and the initial conditions \cite{park2011slow, berglund2000dynamic}. Later, Majumdar \textit{et al.} \cite{majumdar2013transitions} reported that this delay in tipping is independent of the rate of change of control parameter and determined solely by the initial value of the parameter. Recently, Bonciolini \textit{et al.} \cite{bonciolini2018experiments} showed experimentally that the delay in bifurcation increases with rate of change of parameter. After such contradictory observations in literature, Unni \textit{et al.} \cite{unni2019interplay} highlighted the role of interplay between the inherent noise in the system and rate of change of parameter in deciding the tipping point. Even though the delay increases with the rate of change of parameter, noise brings a high variability in the tipping point. Determining the stability margin is difficult for practical systems where stability boundaries are smeared due to \textcolor{black}{this interplay between noise and rate. Devising efficient EWS for abrupt transitions in real systems is of critical importance.} For example, predicting earthquakes, climate changes, and catastrophic events in engineering systems are desirable from social and economic viewpoints. However, predicting such tipping before the occurrence of the event is challenging because the system may not show any indication before the tipping point is reached, especially \textcolor{black}{when there is combined effects of rate and noise}.

Abundant studies discussing quasi-static bifurcations or B-tipping are available in literature. \textcolor{black}{Despite the inherent characteristics of the systems, the dynamics close to the bifurcation point are found to be the same across different systems \cite{schroeder2009fractals}.}
The transitions through various types of bifurcations are related and generic early warning signals exist for catastrophic bifurcations \cite{scheffer2009early}. The phenomenon known as critical slowing down near the bifurcation point gives information about the impending tipping for many types of bifurcations. As the system approaches the critical point, it recovers slowly from the external perturbations. This slow recovery leads to an increase in the memory of the system. The two commonly used early warning indicators that work based on critical slowing down are the lag-1 autocorrelation and the variance of the fluctuations of a system variable. These measures have been proven to predict B-tipping, wherever the tipping is accompanied by a change of stability of the system \cite{scheffer2009early, dakos2008slowing}.

Recently, Wilkat \textit{et al.} \cite{wilkat2019no} showed that there is no evidence of critical slowing down prior to human epileptic seizures. They conjecture that the tipping mechanisms for the human epileptic brain may be a combination of B-tipping, R-tipping and N-tipping and there may be no easily-identifiable EWS for such cases. Most tipping events occurring in nature involve system parameters changing at a constant, varying or undetermined rate along with considerable intensity of noise in the system \cite{tsotsis1988bifurcation, kapila1981arrhenius}. \textcolor{black}{Then, tipping can be influenced} either by noise or by the effect of rate of change of the parameter.
Thus, prediction becomes hard with conventional EWS.
There are many other issues when dealing with rate dependent phenomena. For example, consider the case where the entire transition happens at a very fast rate such that there is not enough data available. Computing precursors in such conditions will be challenging. Even if we get warning, will there be enough time to perform a control action? The outstanding question is: `can we predict transitions in the real systems, considering the \textcolor{black}{combined effects of inherent noise and the rate of change of the parameter?'}

Many studies \textcolor{black}{focused on continuous variation of parameters employs} numerical analysis of standard bifurcation models \cite{bilinsky2018slow, ashwin2017parameter} and limited experimental studies are available \cite{bonciolini2018experiments,scharpf1987experimental, tony2017experimental, pisarchik2014critical}.
In the present study, we choose to work with a prototypical thermoacoustic system (known as a horizontal Rijke tube) exhibiting \textcolor{black}{Hopf bifurcation}, because (i) we observe a catastrophic transition similar to that observed in many practical situations, (ii) we can obtain time series data for a long duration with high sampling frequency, (iii) we can vary the control parameter at different rates and, (iv) we can repeat the experiments at same conditions to verify the observations.

The Rijke tube undergoes a subcritical Hopf bifurcation from a non-oscillatory to an oscillatory state as we vary the control parameter\cite{matveev2003thermoacoustic, juniper2011triggering, gopalakrishnan2015effect}.
The oscillatory state with self-sustained large amplitude periodic oscillations is termed thermoacoustic instability (TAI) in literature \cite{juniper2018sensitivity}.
In a confined thermoacoustic system, the positive feedback between the acoustic field and the unsteady heat release rate manifests as high amplitude acoustic pressure oscillations during TAI. The onset of TAI can cause failure of rocket engines \cite{fisher2009remembering} and damage gas turbine engines \cite{lieuwen2005combustion, fleming1998turbine}. Often, control parameters in practical systems are changed continuously at a finite non-zero rate. Developing EWS for \textcolor{black}{transition to TAI will help to evade such disastrous events}. 
In this paper, we study the \textcolor{black}{effects of rate of change of control parameter on the performance of various EWS, by investigating the variation of warning time provided by EWS with the rate of change of parameter}. Further, we present a mathematical model that captures qualitatively, the key features observed in the experiments.


The rest of the paper is organized as follows. Section II describes the experimental setup. Subsequently, the results and discussions are detailed in Sec. III and the main conclusions are given in Sec. IV. We provide the details of the methodologies for calculating different measures in Appendix A. \textcolor{black}{We show the robustness of EWS with the selection of threshold for warning in Appendix B and the analysis to check for false warnings in Appendix C.}

\section{Experiments}

\begin{figure}
    \centering
    \includegraphics[width=0.5\textwidth]{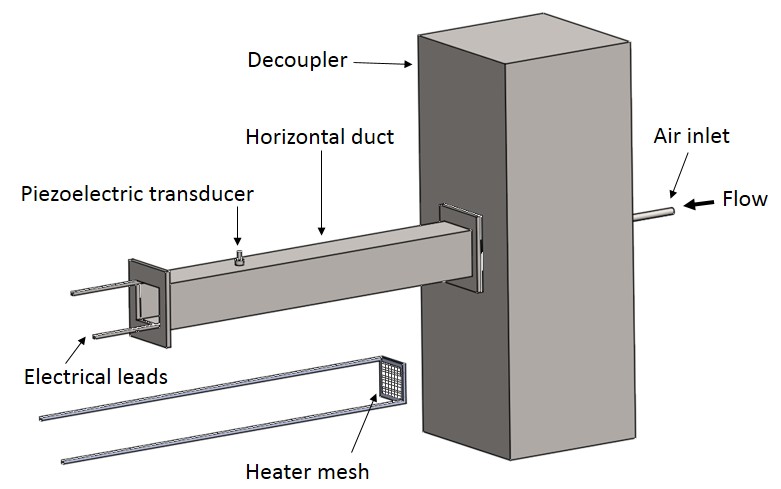}
    \caption{Schematic of the horizontal Rijke tube used for the experiments. It comprises a 1 m long duct, an electrically heated wire mesh and a rectangular chamber called decoupler. The wire mesh is shown separately outside the duct. A piezoelectric transducer is mounted on the duct to acquire the acoustic pressure fluctuations inside the duct.}\label{fig1}
\end{figure}

We perform experiments on a laminar thermoacoustic system known as the horizontal Rijke tube (Fig.~\ref{fig1}). The setup consists of a horizontal duct with a square cross-section which houses an electrically heated wire mesh. Air enters the duct through a rectangular chamber known as the decoupler, which isolates the duct from the upstream fluctuations. The decoupler ensures that the pressure fluctuations at the end connected to it are negligible. The duct is open to the atmosphere at the end away from decoupler. Thus, the pressure at both the ends becomes equal to the atmospheric pressure. This design helps to maintain an acoustically open boundary condition (i.e., acoustic pressure fluctuations, $p^\prime$ = 0) at both the ends. A DC power supply (TDK-Lambda, GEN 8-400, 0-8 V, 0-400 A) is used to provide electric power to heat the wire mesh. 
The mass flow rate of air through the duct is controlled using an electronic mass flow controller (Alicat Scientific, MCR series) with an uncertainty of $\pm$(0.8\% of reading + 0.2\% of full scale). We measure the acoustic pressure fluctuations inside the duct using a piezoelectric sensor (PCB103B02, sensitivity: 217.5 mV/kPa, resolution: 0.2 Pa and uncertainty: 0.15 Pa) at a sampling rate of 10 kHz. A more detailed description of the setup can be found in Gopalakrishnan $\&$ Sujith \cite{gopalakrishnan2015effect}.

In the present study, we control the voltage ($V$) applied across the wire mesh and the current through the mesh changes accordingly.  
Therefore, we estimate the power ($P$) generated in the wire mesh by measuring both the voltage and current. The uncertainty in the measurement of voltage is (0.1 $ V_{rated} +$ 0.1 $V_{measured}$)$\%$ and the uncertainty in the measurement of current is (0.3 $ I_{rated} +$ 0.1 $I_{measured}$)$\%$, where $V_{rated}$ = 8 V and $I_{rated}$ = 400 A. All the other parameters such as the location of the heater mesh (27.5 cm from the decoupler) and the mass flow rate of air (100 SLPM) are kept constant suitably to obtain subcritical Hopf bifurcation.

First, we perform experiments where $V$ is varied in a quasi-steady manner by allowing the system to evolve for a finite time duration at a constant $P$. We let the system reach its asymptotic state and then measure the acoustic pressure fluctuations at different values of $P$. We select the maximum heater power to be less than 1152 W for all the experiments, as the wire mesh may melt and get damaged due to overheating at higher powers. Subsequently, we increase $V$ continuously at different rates and record the pressure signals during the ramp. Here, ramp refers to the continuous increase of the heater power in time. In this paper, we report a linear variation of $V$; \textit{i.e}, the rate ($r = dV/dt$) of change of $V$ is constant ($V = V_0+r\ t$). 
As we have programmed $V$ to vary linearly, $I$ changes accordingly and then the corresponding variation of $P$ is \textcolor{black}{quadratic}. 


\section{Results}

\subsection{\textcolor{black}{Rate-dependent tipping-delay in a thermoacoustic system}}
\begin{figure*}
    \centering
        \includegraphics[width=0.9\textwidth]{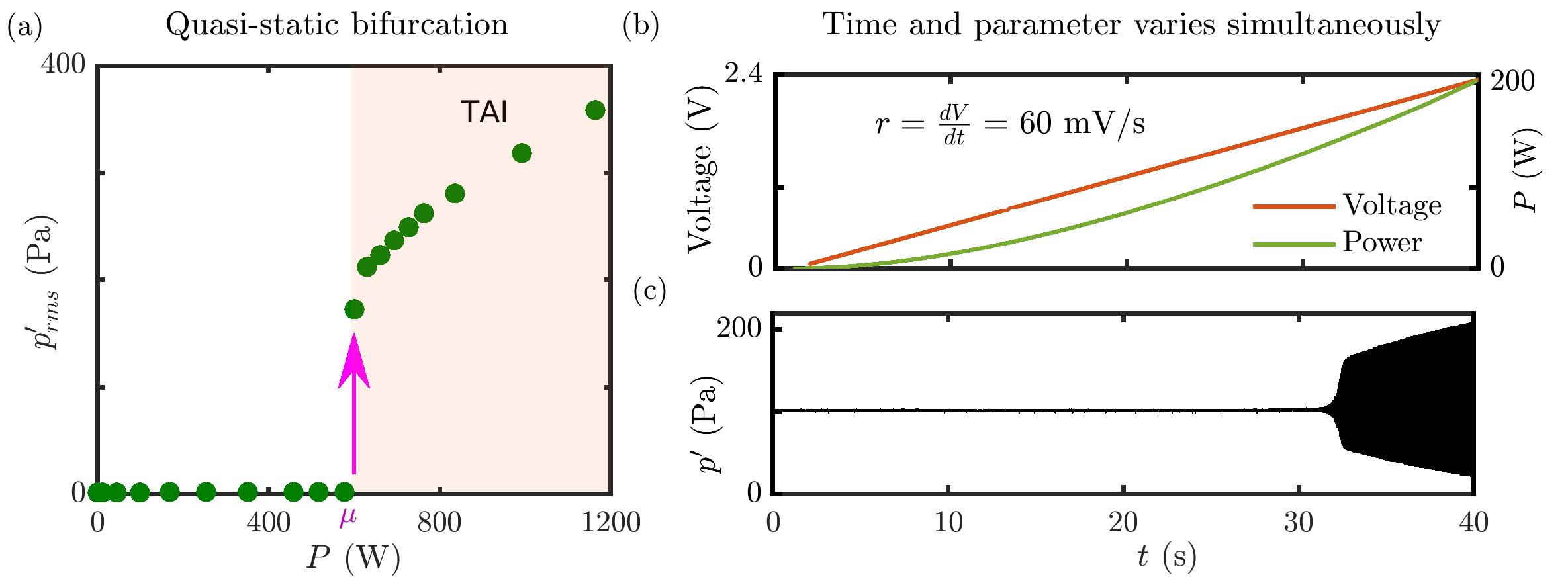}
    \caption{(a) The variation of \textit{rms} of the acoustic pressure fluctuations ($p'_{rms}$) for the quasi-static experiments. $p'_{rms}$ shows an abrupt jump at the heater power $\sim$600 W, denoted as $\mu$. (b) A typical variation of voltage ($V$) at a constant rate of increase (${dV}/{dt}$ = 60 mV/s) and the corresponding variation of heater power ($P$). (c) The pressure signal ($p'$) acquired continuously as the control parameter is varied in time.}
    \label{fig2} 
\end{figure*}

\textcolor{black}{We first conduct quasi-static experiments} to identify the range of parameter values required to capture the transition to high amplitude limit cycle oscillations. We calculate the root mean square (\textit{rms}) value of the acoustic pressure fluctuations ($p'_{rms}$) acquired at different values of heater power ($P$).
Figure~\ref{fig2}a represents the bifurcation diagram showing the variation of $p'_{rms}$ as a function of $P$. We observe that $p'_{rms} \approx 0$ for a range of $P$ corresponding to a quiescent state with amplitude levels comparable to the noise floor ($\sim$ 5 Pa) of the system. At a particular control parameter value, the system transitions to high amplitude limit cycle oscillations. This transition is reflected in $p'_{rms}$ as an abrupt rise and is attributed to subcritical Hopf bifurcation. The parameter value at which this transition occurs, marked as $\mu$ in Fig.~\ref{fig2}a, is known as the Hopf point.
\begin{figure}
    \centering
    \includegraphics[width=0.5\textwidth]{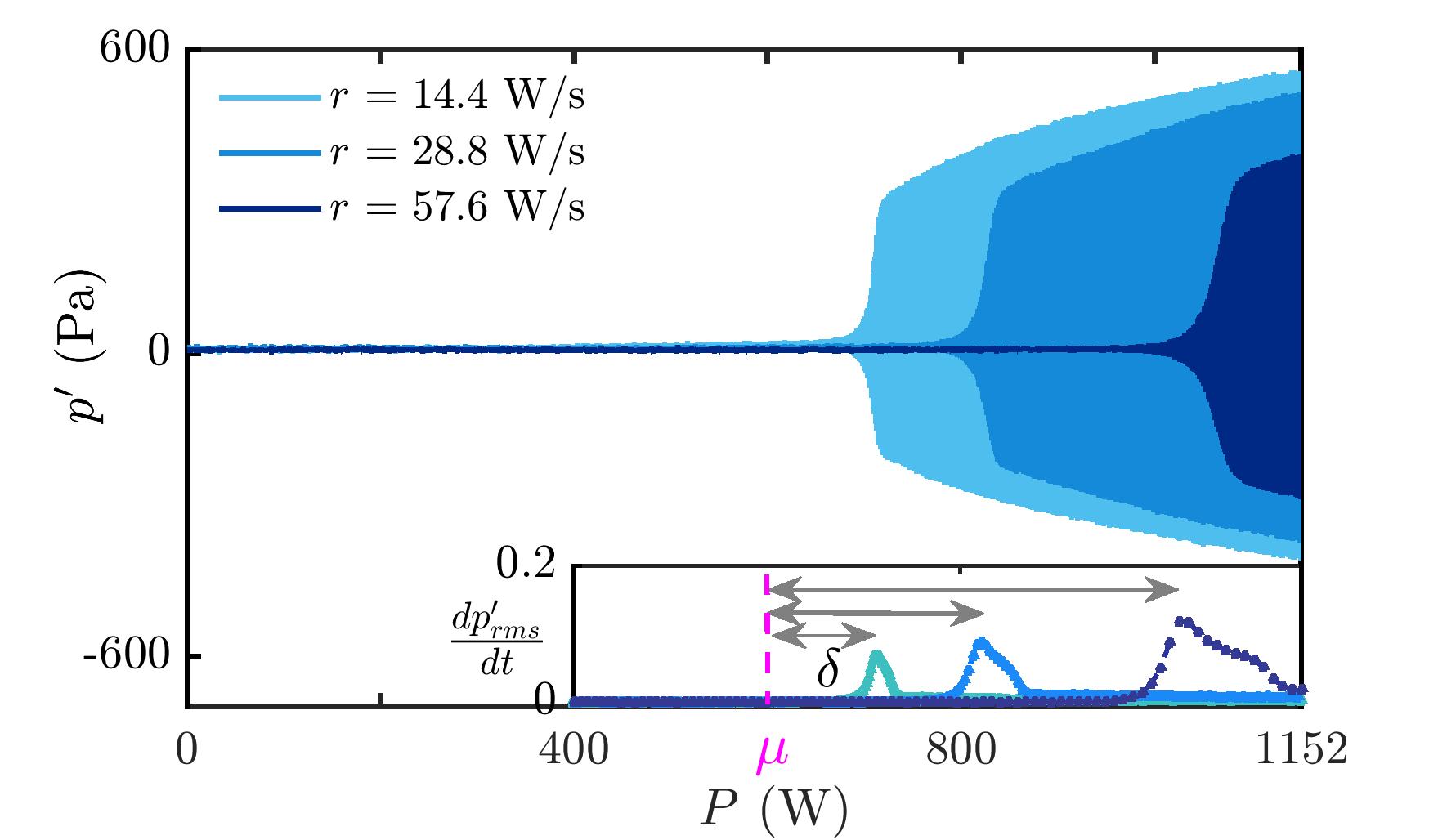}
    \caption{Time evolution of $p'$ as a function of time-varying $P$ for three different rates of change of $V$. The rate of change of ${rms}$ of fluctuations is also shown in the inset to identify the onset of TAI. Here, the maximum rate of change of $p'_{rms}$ is considered as the onset of oscillations. The delay in the transition, $\delta$, is found to increase with an increase in $r$. Here, $p'_{rms}$ is calculated for a moving window of 1 s with an overlap of 0.9 s. }
    \label{fig3}
\end{figure}

As mentioned earlier, we perform experiments with a linear variation of $V$. Therefore, $P$ changes continuously starting from a heater power which is far lower than $\mu$ and increases through the Hopf point to a high value. Here, the variation of $P$ is from 0 to 1152 W in a nonlinear fashion. Thus, throughout this study, we mention the rate of change of voltage with time ($r = {dV}/{dt}$) which is kept constant for each run of the experiment. In Fig.~\ref{fig2}b-c, we plot the typical variation of $V$ and $P$ and the corresponding pressure signal ($p'$) depicting the transition from a quiescent state to high amplitude limit cycle oscillations. 

Figure \ref{fig3} provides the evidence for \textcolor{black}{rate-dependent tipping-delay} as reported in literature \cite{baer1989slow, bonciolini2018experiments, unni2019interplay, scharpf1987experimental}.
We plot the time evolution of $p'$ as a function of time-varying $P$ for three different $r$ (30 mV/s, 60 mV/s, and 120 mV/s) in Fig.~\ref{fig3}. 
It is quite challenging to define a tipping point for the onset of oscillations for \textcolor{black}{dynamic bifurcations}, unlike the quasi-static bifurcations. The difficultly in defining a tipping point is because the oscillations take a finite time to grow, and the parameter would have changed to another value by that time. Hence, we adopt the following method to select the \textcolor{black}{tipping} point. We calculate the rate of change of $p'_{rms}$. The sudden increase in the growth rate of oscillations reflects as a maximum in the rate of change of $p'_{rms}$ \textcolor{black}{(shown in the inset of Fig.~\ref{fig3})}. The delay ($\delta$) in the \textcolor{black}{tipping} is marked from the Hopf point to the maximum of rate of change of $p'_{rms}$ (Fig.~\ref{fig3}). Henceforth, we use this method to define the onset of TAI in this study.

According to bifurcation theory, the system loses its stability at the Hopf point, $\mu$. Due to \textcolor{black}{memory} effects, the system continues to be in the vicinity of the unstable attractor for a finite time. \textcolor{black}{This phenomenon of rate-dependent tipping-delay occurs during slow passage through Hopf bifurcation as described by Baer \textit{et al.}\cite{baer1989slow}}. On comparing the delay ($\delta$) in the onset of TAI for different $r$, we observe that $\delta$ increases with an increase in $r$ (see Fig.~\ref{fig3}), congruent with the observations reported in other systems \cite{baer1989slow, bonciolini2018experiments, unni2019interplay, scharpf1987experimental}. In the case of $r$ = 120 mV/s (the fastest shown in Fig.~\ref{fig3}), we observe a delayed onset of $\sim$470 W from $\mu$. Furthermore, we vary the control parameter at very fast rates, but limiting the values of $P$ to 1152 W so as to not damage the heated wire mesh. In such cases, we do not observe tipping within the duration during which the power is varied; instead, it occurs when we let the system evolve at the final value of $P$, allowing more time. Hence, we confirm that the delay in the tipping increases with rate, by performing experiments at various rates. This trend need not be the same for highly turbulent systems where the inherent fluctuations can perturb the unstable fixed point, causing it to tip towards the stable limit cycle.

\subsection{\textcolor{black}{EWS for critical transitions}}

\begin{figure*}
    \centering
    \includegraphics[width=0.8\textwidth]{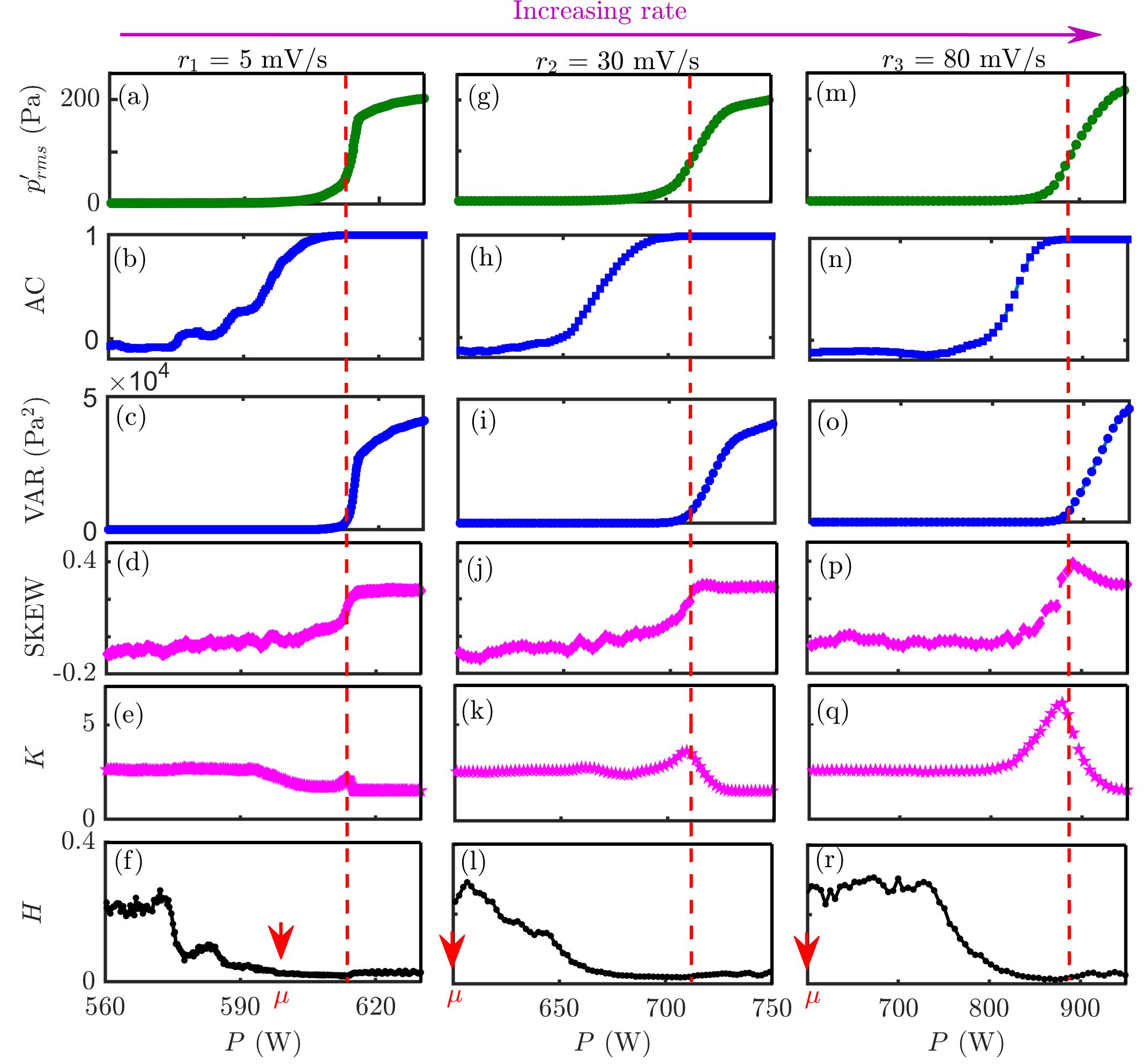}
    \caption{Variation of $p'_{rms}$, lag-1 autocorrelation (AC), variance (VAR), skewness (SKEW), kurtosis ($K$) and Hurst exponent ($H$) \textcolor{black}{during the transition to TAI in} Rijke tube. Each column corresponds to the results for a particular rate ($r$) of change of $V$ \textcolor{black}{((a)-(f): $r =$  5mV/s, (g)-(l): $r =$  30mV/s, (m)-(r): - $r =$  80mV/s)}. The onset of TAI is marked with a red colour dotted line and quasi-static Hopf point $\mu$ is marked on the \textit{x}-axis. $p'_{rms}$, and VAR starts to increase almost at the onset of TAI, whereas, $K$ and SKEW detect the tipping slightly before $p'_{rms}$ increases. In contrast, AC and $H$ give early warning well-before the transition to TAI for all the cases shown here.}
    \label{fig4}
\end{figure*}
We compute several EWS for the pressure signal acquired continuously during the ramp for the experiments performed at different rates of change of control parameter. The following measures are considered in this study: $p'_{rms}$, lag-1 autocorrelation (AC), variance (VAR), skewness (SKEW), kurtosis ($K$), and Hurst exponent ($H$). 

Autocorrelation calculates the correlation of a signal with its delayed copy as a function of the delay. Lag-1 autocorrelation is the correlation between values in the signal that are one time step apart. In this study, we refer to lag-1 autocorrelation as AC. Variance (VAR) is the expectation of the squared deviation from the mean. For critical transitions, both AC and VAR increase based on the phenomenon of critical slowing down. Systems approaching a transition to a new stable state, where the current stable state becomes unstable due to change in the control parameter, show slow response to external perturbations. The phenomenon of slow recovery rate to the external perturbations close to a critical transition is known as critical slowing down \cite{strogatz1994nonlinear} (CSD). This slowing down leads to an increase in autocorrelation and variance of fluctuations \cite{scheffer2009early, dakos2012robustness}. 

When a system is driven gradually closer to a critical transition, the increase in the autocorrelation occurs much before the actual transition. Similarly, the impact of perturbations do not decay fast, and their accumulating effect increases the variance of fluctuations. CSD results in increased short term memory of the system, and hence autocorrelation at low lags would increase. \textcolor{black}{CSD is observed in realistic models of spatially complex systems \cite{lenton2009using} as well as in simple models \cite{dakos2008slowing} and has been used as EWS for critical transitions\cite{scheffer2009early}.}

Skewness (SKEW) and kurtosis ($K$) are not directly connected to critical slowing down. SKEW is a measure of the symmetry of the probability distribution of the data about its mean; \textit{i.e}., whether the distribution is biased towards one side over the other. A distribution is said to be negatively skewed (or left skewed) when a majority of the data falls to the right of the mean. On the other hand, a distribution is positively skewed (or right skewed) when more data is concentrated to the left of the mean. SKEW is zero for a normal distribution. Generally, asymmetry of fluctuations may increase (SKEW increases) on approaching a catastrophic bifurcation, as the potential landscape near the transition would be less steep on one side of the equilibrium than the other \cite{guttal2008changing, scheffer2009early}. Kurtosis ($K$) gives information about whether the distribution has heavy tails, or is more centred. The kurtosis of a normal distribution is 3, and a higher $K$ indicates more outliers in the data. Very close to the tipping point, a comparatively longer distribution with fatter tails results in a high value of kurtosis, $K>3$.  A detailed procedure for the computation of these EWS is given in Appendix A-3.

Apart from these conventional measures, we investigate the fractal characteristics of the data close to tipping \cite{nair2014multifractality,sujith2020complex}. We use a measure known as the Hurst exponent ($H$), which is related to the fractal dimension ($D$) of the time series \cite{mandelbrot1983fractal,hurst1951long} as $H=2-D$. Fractal objects exhibit self-similar features at various scales of magnification; therefore, measures such as length, area, and volume are dependent on the scale of measurement. For a fractal time series, the scaling of \textcolor{black}{the rms of the standard deviation of fluctuations} with the length of the data segment gives $H$. For non-fractal objects such as sinusoidal signals, $H \approx 0$ as there is no scaling with the data length or the scale of the measurement. Unlike mathematical fractal objects which possess self-similarity across a wide range of scales extending up to infinity, real fractal objects appear to be self-similar only over a limited range of scales; one cannot keep on zooming in indefinitely to see the same structure. Thus, the time scales for calculation \textcolor{black}{of $H$} need to be selected carefully. As the system has preferred time scales (corresponding to the natural frequency) in the present study, selecting time scales less than one acoustic cycle may not capture the periodicity present in the data and more than 4 acoustic cycles may average out the fluctuations. Hence, we choose 2 to 4 acoustic cycles for the calculation of Hurst exponent \cite{nair2014multifractality}.
We calculate $H$ following the algorithm of Multifractal Detrended Fluctuation Analysis (MFDFA)\cite{kantelhardt2002multifractal}, which is described in detail in Appendix A-4.

$H$ is a measure of persistence or correlation of a signal. If an increase in the value is more likely to be followed by another increase in value, then the signal is called persistent. A persistent or a positively correlated signal has $H > 0.5$, an anti-persistent or negatively correlated signal (an increase in value is mostly followed by a decrease in value, or vice versa) have $H < 0.5$ and uncorrelated white noise has $H = 0.5$. 
Fractal analysis has found a variety of applications in life sciences, engineering, econophysics, and geophysics \cite{ivanov1999multifractality, hu2004endogenous, grech2004can, vandewalle1997coherent, grech2008local, alvarez2008time, matos2008time,  domino2011use, suyal2009nonlinear, kilcik2009nonlinear, nair2014multifractality,unni2015multifractal, gotoda2012characterization}. For instance, it has been used to distinguish healthy patients from patients with heart failures \cite{ivanov1999multifractality, havlin1999application}. Similarly, the variations in the $H$ of geoelectric and seismic fluctuations provide indicators for earthquakes \cite{telesca2001new}. In econophysics, Qian $et\ al.$ \cite{qian2004hurst} used $H$ as a measure of financial market predictability. 
In the present study, we use $H$ to predict \textcolor{black}{critical transitions} in a thermoacoustic system.

\textcolor{black}{To study the effect of rate of change of parameter on the performance of the aforementioned EWS}, we plot the variation of these measures as a function of $P$ for three representative cases of $r$: 5 mV/s, 30 mV/s and 80 mV/s in the first, second and third column, respectively (Fig.~\ref{fig4}). We compute the measures for a moving window of size 1 s with an overlap of 0.9 s. The choice of this particular window size is to ensure that at least 100 cycles of oscillations are covered in a window. The Hopf point ($\mu$) is marked at 600 W with a red coloured arrow for reference, even though the tipping occurs after a delay. 
$p'_{rms}$ indicates the growth in the amplitude of oscillations during the transition. VAR detects the transition almost at the same $P$ where the amplitude rises which is reflected as a steep increase in the $p'_{rms}$ at the onset of TAI, while SKEW and $K$ seem to perform slightly better than $p'_{rms}$ and VAR. Initially, we observe small negative SKEW values indicating a slightly left/negatively skewed distribution, and during the transition, it shifts to a slightly right/positively skewed distribution. Nevertheless, there is no significant change in SKEW during the transition, since $-0.5<$ SKEW $< 0.5 $ is generally considered as symmetric distribution. Thus, the distribution has not changed in terms of skewness, and we cannot consider the change in SKEW as a precursor for the tipping. Besides, kurtosis has a value $\sim$3 pertaining to a normal distribution during the quiescent state. After the transition to TAI, kurtosis reduces to a value lower than 3. It appears that in all the cases, there is a local maxima for the kurtosis at the \textcolor{black}{onset of TAI}. However, we observe a drop in the kurtosis for the first case (Fig.~\ref{fig4}e), prior to the tipping point. Close to the tipping, the variation of kurtosis is not consistent for different rates: it reduces for the first case (Fig.~\ref{fig4}e) and increases for the other cases (Fig.~\ref{fig4}k,q). \textcolor{black}{Hence, we do not recommend $K$ as a good EWS for critical transitions in practical systems.} 

\begin{figure}[t]
    \centering
    \includegraphics[width=0.47\textwidth]{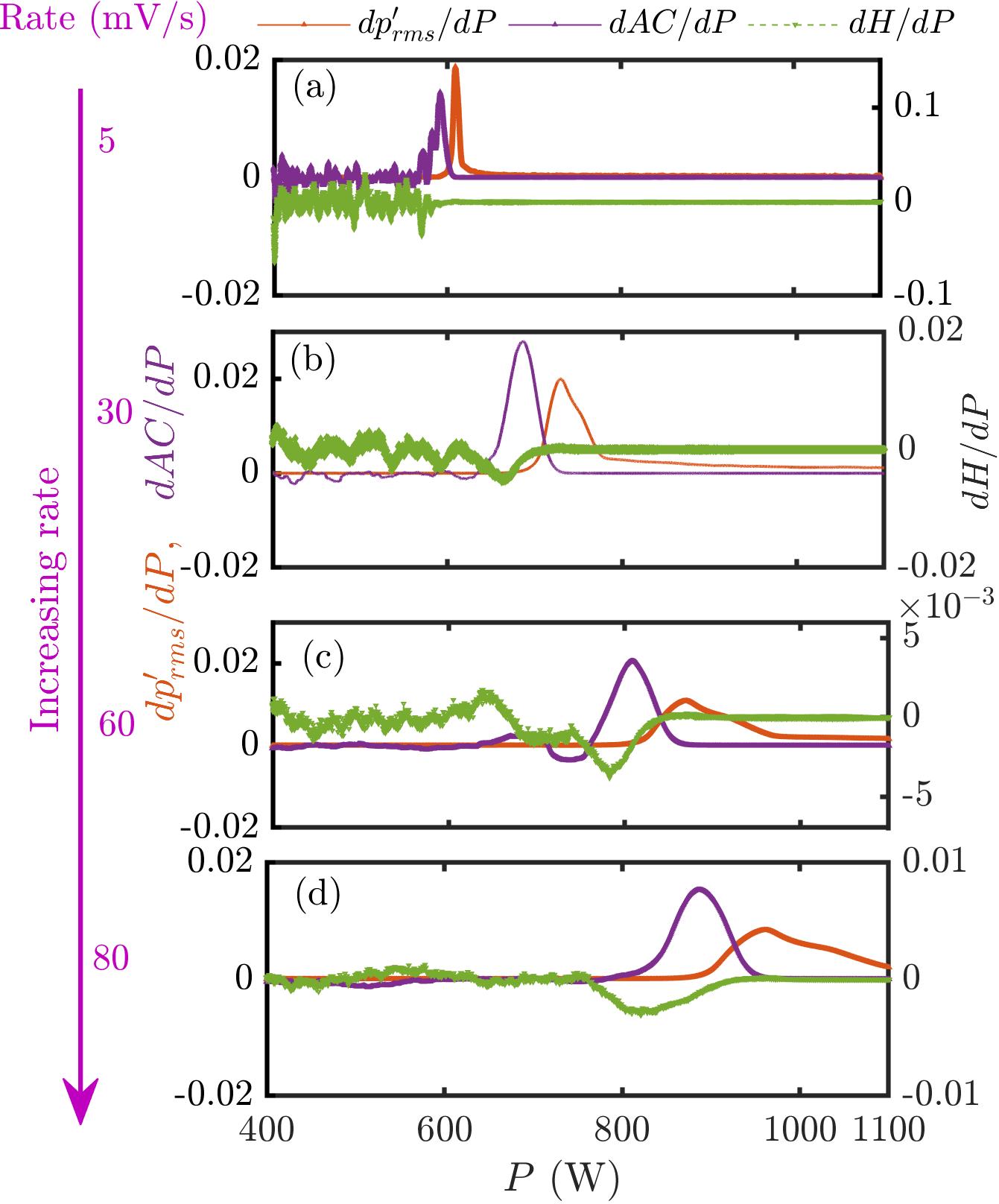}
    \caption{Rate of change of $p'_{rms}$, lag-1 autocorrelation (AC), and Hurst exponent ($H$) with $P$ is plotted for different rates of change of $V$. The maximum rate of change of AC and $H$ occurs much before the maximum growth of $p'_{rms}$. As the rate of change of input voltage with time (${dV}/{dt}$) increases  \textcolor{black}{(from (a)-(d))}, the delay  \textcolor{black}{(in terms of $P$)} in growth of amplitude increases. }
    \label{fig_rateEWS}
\end{figure}

Among these conventional measures, lag-1 autocorrelation (AC) appears to be the best EWS for this type of transition. An initial quiescent state results in near zero AC due to the low amplitude uncorrelated noise. AC increases as periodicity increases and approaches 1 for limit cycle oscillations. It is clear from these experiments that AC is a more robust early warning measure compared \textcolor{black}{to} $p'_{rms}$, VAR, SKEW and $K$. Earlier studies \cite{gopalakrishnan2016early, 6848858} have reported that autocorrelation in the presence of fluctuations is a less effective precursor compared to variance. Actually, variance starts to build up gradually long before the transition, but it increases rapidly \textcolor{black}{only} at the tipping point. \textcolor{black}{In fact, variance is the square of $p'_{rms}$}. As we do not know the amplitude of oscillations in the final state, it is difficult to rely on variance to determine when the transition will take place. In contrast, the value of AC is bounded between 0 and 1, and we know  \textcolor{black}{how close we are to the tipping from the value of AC}.

The fractal based measure, $H$ fluctuates between the values 0.2 to 0.5 for the fixed point state. Even though the flow field is laminar, there is noise present in the fixed point state originating from different sources such as the compressor or electronic noises. Therefore, the fixed point state with low amplitude aperiodic noisy fluctuations would give $H$ values between 0.2 and 0.5. During the transition to the state of limit cycle oscillations, $H$ approaches zero. $H$ captures the periodicity (or the loss of fractal nature) in the data even if the amplitudes are very low. 
We start observing an emergence of periodicity as we approach the tipping point.
The inherent noisy fluctuations perturb the system from the stable fixed point. These noise-induced oscillations occur at frequencies centred around the natural frequency of the system and the oscillations decay in time. These fluctuations contain very low amplitude bursts of periodicity which has oscillations around the natural frequency of the system\cite{wiesenfeld1985noisy}. Capturing this slight periodicity in the system variables close to the transition, $H$ starts to decrease towards 0 well-before the \textit{rms} of fluctuations grows. For very slow rates, $H$ tends to decrease before $\mu$ (Fig.~\ref{fig4}f). For relatively faster rates, the tipping is delayed significantly from $\mu$; nevertheless, $H$ forewarns the tipping well-before the rise in $p'_{rms}$ (Fig.~\ref{fig4}l,r). 

We repeat the computations of all EWS for data acquired at many different rates of change of $V$. The results shown in Fig.~\ref{fig4} are for three representative cases from this collection of data. A similar inference is obtained for even faster rates of change of voltage up to 240 mV/s. In all the cases, AC and $H$ prove to be the better measures to forewarn an impending TAI compared to other measures such as VAR, SKEW and $K$, and both AC and $H$ have comparable effectiveness.

In summary,  there are two things happening on approaching the transition; one is the growth of the amplitude of oscillations due to Hopf bifurcation, and the second is the increase in temporal correlation. Out of all the EWS discussed here, $rms$ and variance capture the growth of amplitude of oscillations;  skewness and kurtosis detect the changing distribution of data during the transition; lag-1 autocorrelation determines the increasing correlation between consecutive time instants; $H$ looks at the increasing correlation as well as the emergence of periodicity. AC and $H$ are capturing features that are different from the features captured by $p'_{rms}$, VAR, SKEW and $K$. Variance, skewness and kurtosis do not change if we were to shuffle the data randomly. However, AC and $H$ will change upon shuffling the data. Hence, AC and $H$ are capturing the temporal correlations present in the data more than just the \textcolor{black}{statistical characteristics of the data.} From Fig.~\ref{fig_rateEWS}, it is clear that the increase in correlation occurs prior to the growth of amplitude. Therefore, $AC$ and $H$ are able to predict the tipping well in advance. For faster rates of change of control parameter, growth of correlation occurs at a much lower parameter value than the growth of amplitude in the signal.

\subsection{Variation of warning time with rate}

Our analysis shows that AC and $H$ are \textcolor{black}{effective EWS for critical transitions} in the considered thermoacoustic system. Next, we compare the early warning time for different rates using AC and $H$. Till now, we were focused on the warning in the parameter space, \textit{i.e.}, at what parameter value we can predict, compared to the tipping point. Ultimately, EWS need to be compared across both temporal domain and parameter space.
\begin{figure*}
\centering
    \includegraphics[width=0.98\textwidth]{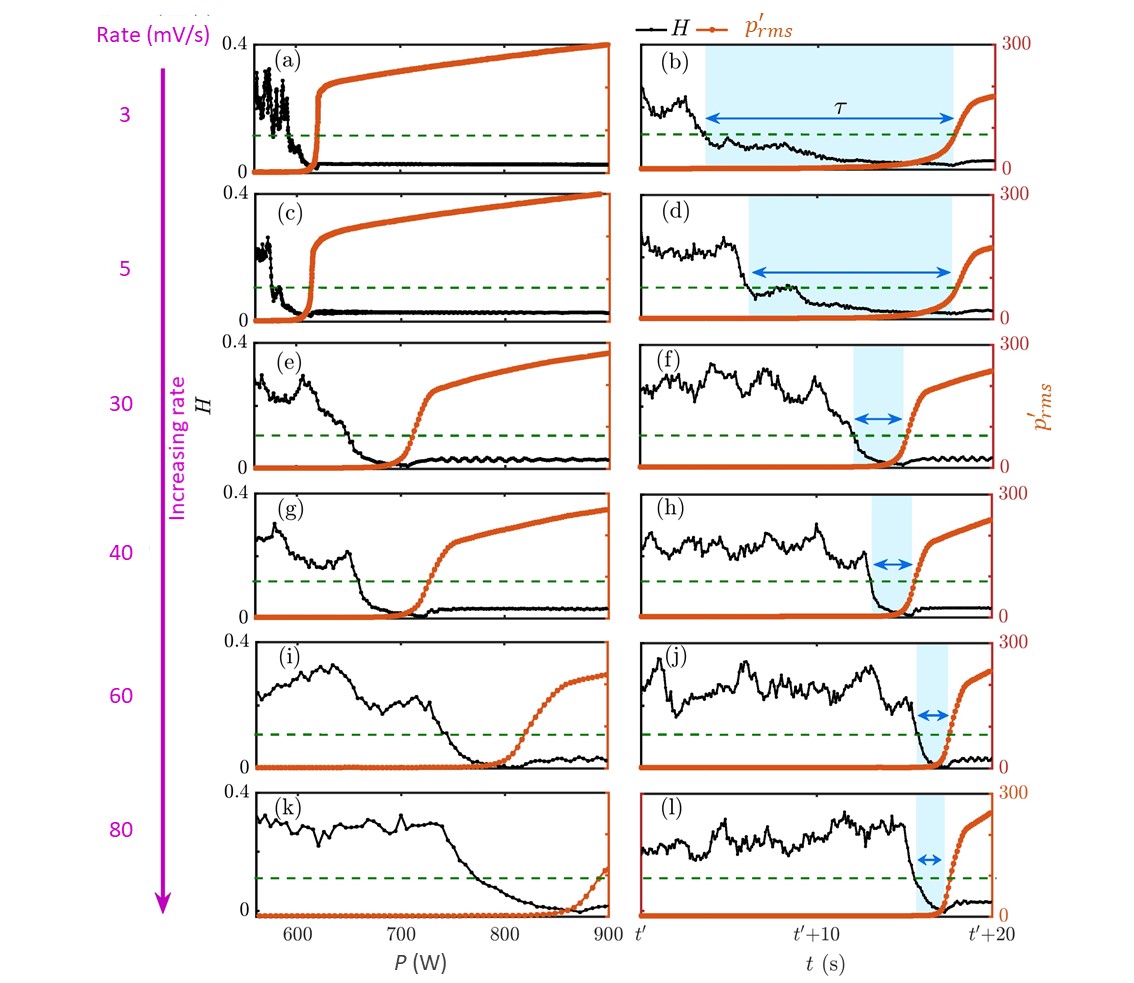}
    \caption{Variation of $p'_{rms}$ and $H$ evaluated for data acquired at different rates of change of $V$ is plotted as a function of the $P$ (left column -  \textcolor{black}{a,c,e,g,i,k}) and time (right column - \textcolor{black}{b,d,f,h,j,l}). Rate of change of parameter increases from top to bottom. For all these rates, we are able to predict the tipping using $H$, before $p'_{rms}$ starts to increase. The green dotted line represents to the threshold value of $H$ = 0.1. \textcolor{black}{The warning time before the tipping (marked with blue colour) is the difference between the time at which $H$ crosses the threshold and the time at which maximum rate of change of $p'_{rms}$ is observed.}  } 
    \label{fig6}
\end{figure*}

\begin{figure*}
    \centering
    \includegraphics[width=0.87\textwidth]{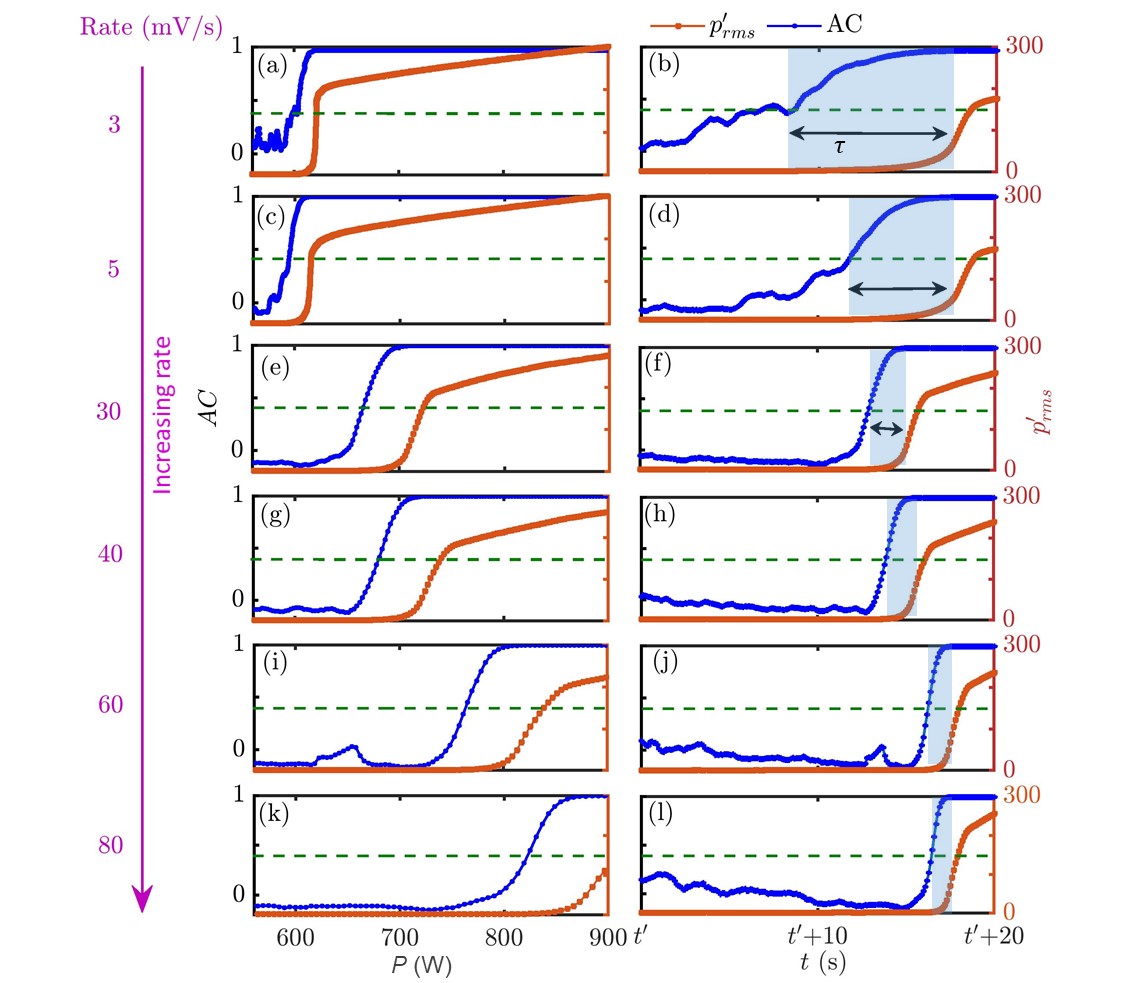}
    \caption{Variation of $p'_{rms}$ and AC is plotted as a function of control parameter, $P$, (left column - \textcolor{black}{a,c,e,g,i,k}) and time, $t$, (right column - \textcolor{black}{b,d,f,h,j,l}) evaluated for data acquired at different rates of change of $P$. Rate of change of control parameter increases from top to bottom. For all the transitions shown, AC provides an early warning before $p'_{rms}$ starts to rise. The green dotted line represents to the threshold value of $AC$. \textcolor{black}{The warning time, $\tau$, (marked with blue colour) is the difference between the time at which AC crosses the threshold and the time at which maximum rate of change of $p'_{rms}$ is observed.}} 
    \label{fig7}
\end{figure*}

\begin{figure*}
    \centering
    \includegraphics[width=0.8\textwidth]{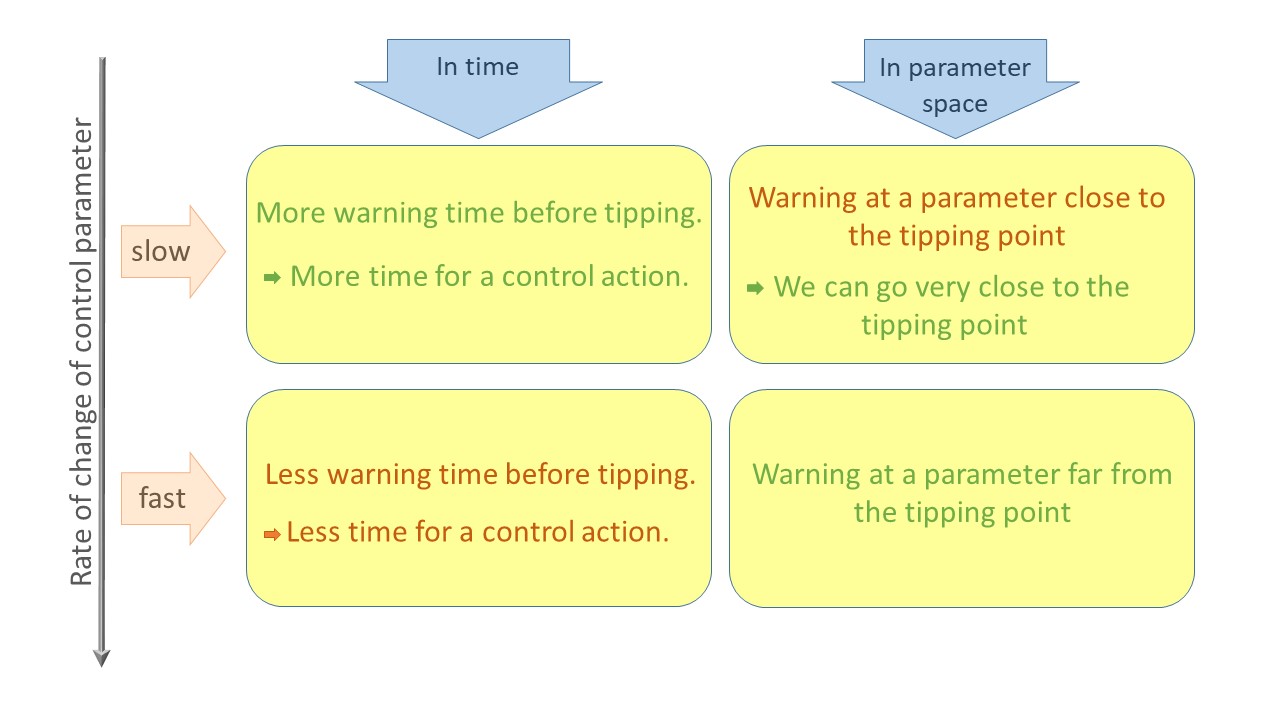}
    \caption{The effect of rate of change of control parameter on EWS for \textcolor{black}{critical transitions}} 
    \label{fig_quadchart}
\end{figure*}

Figure~\ref{fig6}-\ref{fig7} show the variation of $H$ and $AC$ as a function of the control parameter (left column) and time (right column), \textcolor{black}{for different values of $r$}. Variation of $p^\prime_{rms}$ is plotted on the right axis for comparison. A window of 20 s time interval is shown (right column) for all the plots, for the sake of comparison. The time $t'$ is selected as a time instant before the tipping point within a window of 20 s. The warning time is marked as $\tau$ (blue shaded region in Figs.~\ref{fig6}-\ref{fig7}).
We calculate $\tau$ by selecting a threshold in AC and $H$ so that, when it crosses the threshold we are warned of an impending tipping. In the present case, we select AC = 0.4 and $H$ = 0.1 as the threshold, because $H$ never reduces to 0.1 unless the system is proceeding towards an impending tipping. Similarly, AC also has fluctuations, but, AC = 0.4 is out of bounds for the quiescent state. The green dotted line in Figs.~\ref{fig6}-\ref{fig7} corresponds to the selected threshold value of $H$ = 0.1 or AC = 0.4.
The time between $H$ = 0.1 or AC = 0.4 and the onset of TAI (corresponds to the maximum rate of change of $p^\prime_{rms}$) is considered as $\tau$. This choice of threshold for AC and $H$ are not unique, and any other slightly different threshold also will work equally well.
Before the amplitude rises steeply, we have more warning time ($\tau$) in the case of a relatively slower rates shown in Fig.~\ref{fig6}b,d.

In terms of the parameter, we get warning at a parameter value well ahead of the tipping point for faster rates (Figs.~\ref{fig6}k,l and \ref{fig7}k,l). However, the time needed to reach the tipping point is really short, as the rate of change of $P$ is fast. Hence, we have relatively lesser time to implement control actions for faster rates (Figs.~\ref{fig6}k,l and \ref{fig7}k,l). For slower rates, we have more time for control, but we are close to the tipping point in the parameter space (Fig.~\ref{fig6}a,b and \ref{fig7}a,b). 
As we are going closer to the tipping point in case of slower rates, there is a possibility of N-tipping if external noise or inherent fluctuations of significant magnitude are present. 
The analysis of warning time using AC provides a similar inference, wherein slower rates result in more warning time and faster rates give more early warning in terms of the control parameter (Fig.~\ref{fig7}). These findings are summarized in Fig.~\ref{fig_quadchart}.

\begin{figure}[h]
    \centering
    \includegraphics[width=0.5\textwidth]{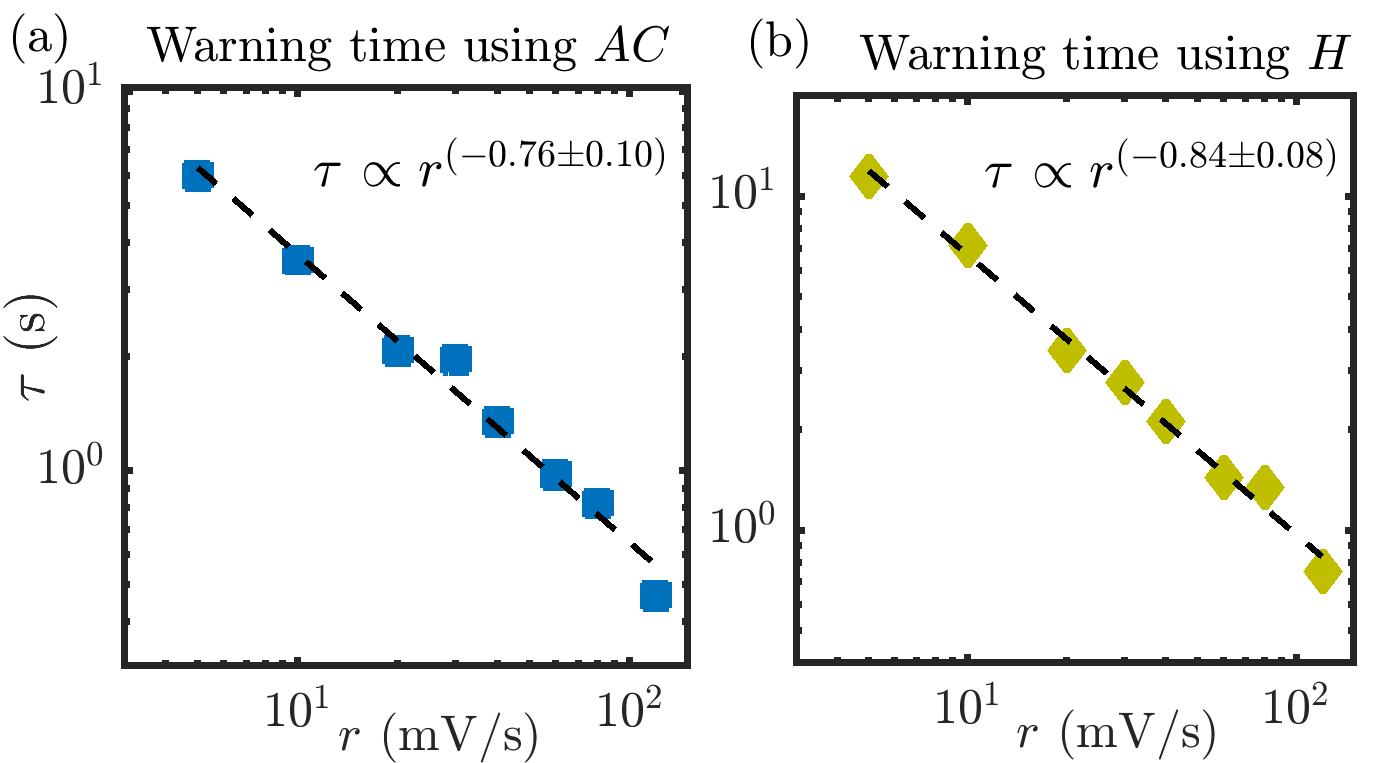}    
    \caption{Warning time ($\tau$) obtained from (a) AC and (b) $H$ are plotted as a function of $r$ in logarithmic scale. (b)  The value of $\tau$ decreases with $r$ following an inverse power law relation as $\tau \propto r^{n}$ with $n$ = -0.76 $\pm$ 0.10 and $n$ = -0.84 $\pm$ 0.08 for AC and $H$, respectively. The threshold for obtaining warning is 0.4 for AC and 0.1 for $H$.}
    \label{fig9}
\end{figure}

Next, we examine how the warning time ($\tau$) varies with the rate of change of voltage with time ($r$ = ${dV}/{dt}$). We calculate $\tau$ by selecting a threshold, as mentioned in the preceding paragraph. As we increase the rate, $\tau$ decreases drastically (see Fig.~\ref{fig9}). We observe an inverse power law relation between the warning time and the rate of change of parameter as $\tau \propto r^{n}$ with $n$ = - 0.76 $\pm$ 0.10 and $n$ = -0.84 $\pm$ 0.08 for AC and $H$, respectively. The uncertainty in fitting is estimated with 95$\%$ confidence. The scaling with a constant exponent holds for a range thresholds for AC (threshold: 0.2-0.8) and $H$ (threshold: 0.15-0.06). The scaling using different thresholds are shown in Appendix B.

\subsection{\textcolor{black}{Model for noisy Hopf bifurcation with continuous variation of parameter}}
In the present study, we use the model of a nonlinear oscillator with additive noise exhibiting subcritical Hopf bifurcation \cite{noiray2017linear}.
\begin{equation}
\ddot{\eta}+\alpha \dot{\eta}+\omega^{2} \eta=\dot{\eta}\left(\beta+K \eta^{2}-\gamma \eta^{4}\right)+\mathcal{N}_1,
\label{eq7}
\end{equation}
\begin{equation*}
\textcolor{black}{\frac{d\alpha}{dt} = 2rt,}
\end{equation*}
where $\eta$ and $\dot{\eta}$ are the state variables, $\alpha$ and $\beta$ are linear damping and driving respectively and $\omega$ is the natural frequency. We use additive Gaussian white noise $ \mathcal{N}_1$ of intensity $\Gamma_1$ with an autocorrelation $<\mathcal{N}_1 \mathcal{N}_{1\ \tau}> = \Gamma_1^2 \delta (\tau)$. Following Noiray \cite{noiray2017linear}, the values of the parameters $\omega$, $\beta$, $\gamma$ and $K$ are kept constant as follows: $\omega$ = 2$\pi \times$ 120 rad/s, $\beta$ = 50 rad/s, $\gamma$ = 0.7, and $K$ = 9. We select a low value of $\Gamma_1 = \sqrt{10^{-1}}$ as we are modelling a laminar system. 
Here, the linear damping ($\alpha$) can be considered as the bifurcation parameter analogous to the control parameter in the experiment. $\alpha$ is reduced from 200 rad/s to -50 rad/s to capture the subcritical Hopf bifurcation from a state of stable fixed point to limit cycle oscillations.

\begin{figure}[h!]
    \centering
    \includegraphics[width=0.48\textwidth]{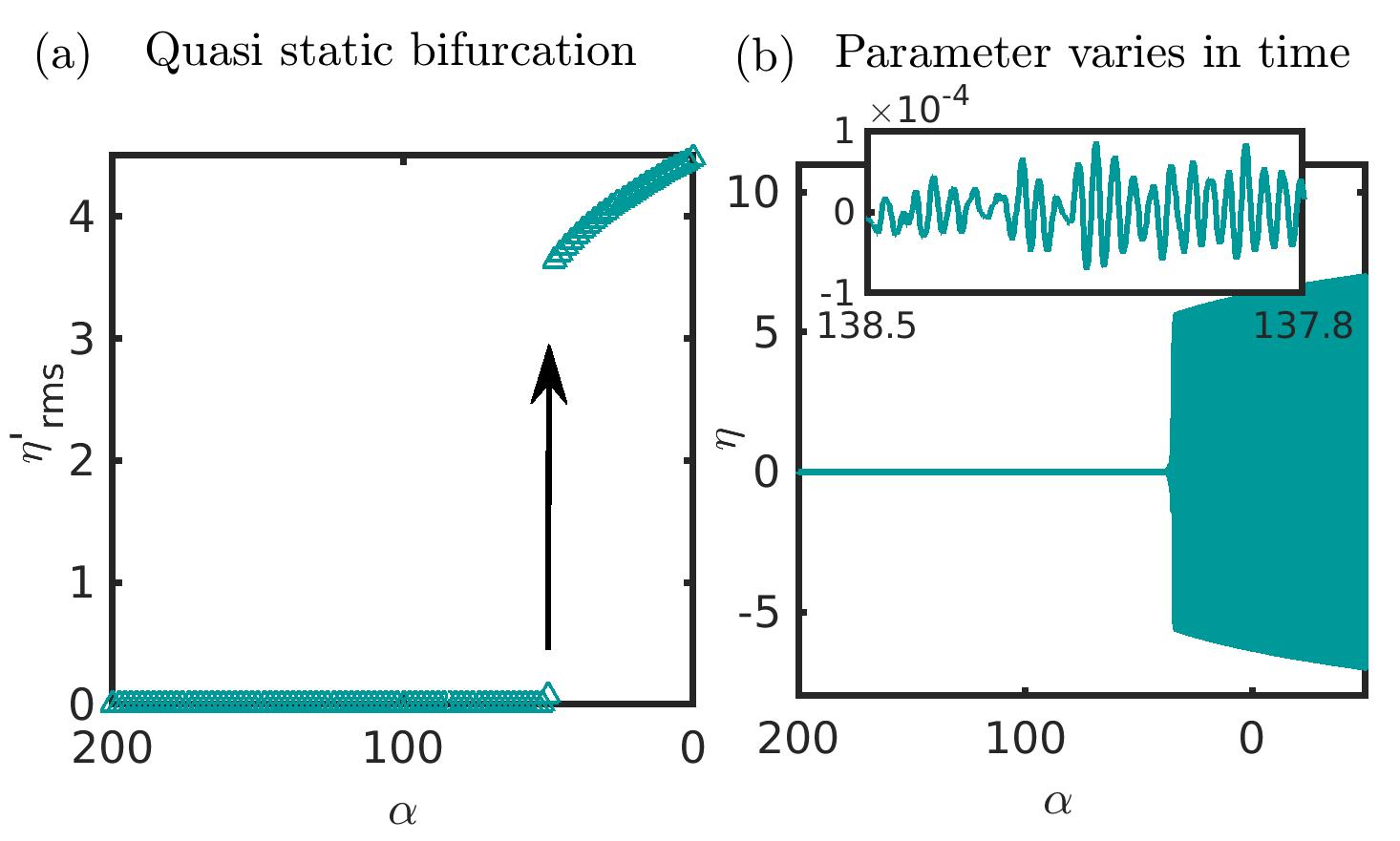}
    \caption{ (a) Bifurcation diagram showing the variation of  $\eta^{\prime}_{rms}$ as a function of linear damping, $\alpha$. The Hopf point ($\mu$) is at $\alpha$ = 50 rad/s, where $\alpha = \beta$. (b) The time series of $\eta$ obtained by solving Eq.~\ref{eq7} for a continuous variation of $\alpha$ is shown along with a zoomed view of the bursts occurring during the fixed point state. }
    \label{fig_model_quasi}
\end{figure}

In order to study the effect of rate of change of parameter in the model, we vary $\alpha$ continuously. In experiments, the heater power is changing nonlinearly even though we vary $V$ linearly. To replicate a similar variation of the control parameter, we change $\alpha$ as $\alpha = \alpha_0 +r \ t^2$ in the model. Initially, the linear damping $\alpha$ is high (200 rad/s) to obtain a fixed point. Upon decreasing $\alpha$, once $\alpha= \beta$, the fixed point becomes unstable. However, there has to be a non-zero perturbation for the system to escape from the fixed point. The additive white noise term ($ \mathcal{N}_1$) in Eq.~\ref{eq7} constantly perturbs the system from the fixed point and helps to jump to the limit cycle state. In experiments, even though the system is laminar, low amplitude noisy fluctuations are always present in the flow field.

Although the above model captures the dynamic transitions qualitatively, the initial fixed point state appears to have bursts with a high level of periodicity (Fig.~\ref{fig_model_quasi}b), unlike the noisy aperiodic fluctuations observed in the experiments. Notably, the fixed point in Hopf bifurcation under the influence of noise contains fluctuations with bursts of periodicity with shallow peaks near the fundamental frequency of the system \cite{wiesenfeld1985noisy, fujisaka1989correlation}. However, the experimental data appears to be more aperiodic. In experiments, the base state with constant airflow and zero heater power itself generates low amplitude aperiodic pressure fluctuations which will also get measured along with the system dynamics. This inherent fluctuations and the measurement noise involved in the dynamics could be modelled by adding Gaussian white noise ($\mathcal{N}_2$) with intensity $\Gamma_2$ to the time series of $\eta (t)$ obtained by solving the model.
\begin{equation}
    \eta (t) = \eta (t) + \mathcal{N}_2(t).
\end{equation}

We choose a noise intensity ratio, $\Gamma_2/\Gamma_1$ = 0.03, to replicate the experimental data qualitatively. A very low value of $\Gamma_2/\Gamma_1$ would result in high values of AC and low values of $H$ due to the presence of low amplitude bursty oscillations during the noisy fixed point state. As we gradually increase the intensity of external noise (\textit{i.e.,} increasing $\Gamma_2/\Gamma_1$), the values of $H$ increases and AC decreases. Then, the values during the transition also matches with those obtained from experiments. However, if we add external noise more than necessary, it would suppress all the underlying dynamics and produce AC = 0 and $H = 0.5$ due to the white noise characteristics. The addition of external noise to the output with a particular ratio of intensity makes the signal more aperiodic and match the characteristics of experimental data.

\begin{figure*}
    \centering
    \includegraphics[width=0.8\textwidth]{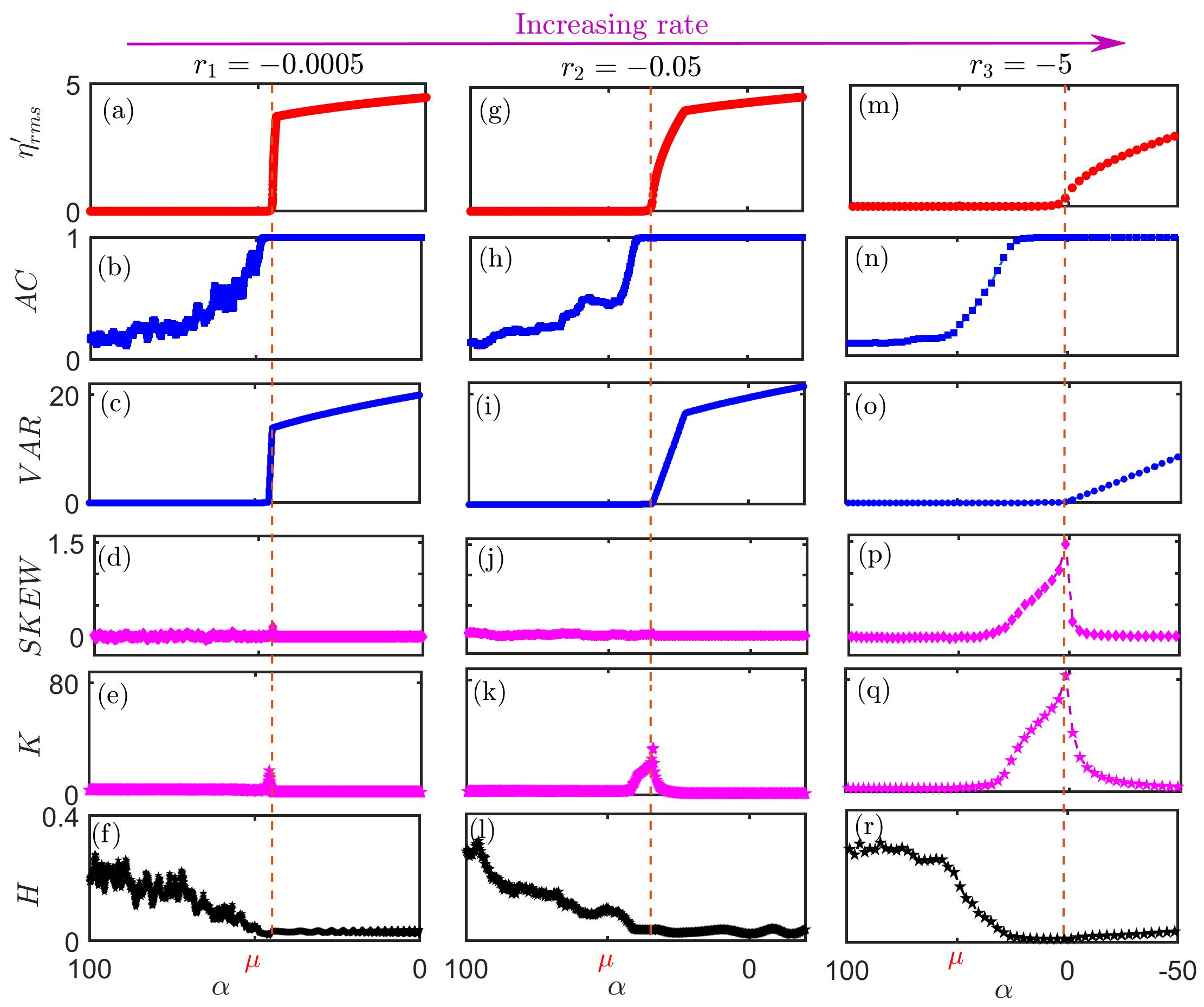}
    \caption{Variation of $\eta'_{rms}$, lag-1 autocorrelation (AC), variance (VAR), skewness (SKEW), kurtosis ($K$) and Hurst exponent ($H$) with the \textcolor{black}{parameter $\alpha$ during the transition to limit cycle oscillations} in the model. Each column corresponds to the results for a particular rate of change of $\alpha$ \textcolor{black}{((a)-(f): $r =$ -0.0005, (g)-(l): $r =$ -0.05, (m)-(r): - $r =$ -5)}. The point of the maximum rate of change of $\eta'_{rms}$ is marked with a red dotted line. For higher values of $r$, kurtosis and skewness detect the tipping before $\eta'_{rms}$ and VAR. Here, AC and $H$ consistently give early warning well-before the transition to limit cycle oscillations for all the rates.}
    \label{fig11}
\end{figure*}

Now, with this model, all the previously discussed EWS are computed for data generated with different rates of change of $\alpha$ is shown in Fig.~\ref{fig11}. For all the measures, we observe a similar performance as that observed for the experiments. Again, AC and $H$ provide a better warning for an impending TAI compared to the other measures. For faster rates of change of parameter, SKEW and $K$ predict the tipping slightly better for the model when compared to the experiments. For slower rates of change of parameter, the transition is more abrupt in terms of the control parameter and all measures except AC and $H$ detect this very close to the tipping point. In contrast, the growth of amplitude of fluctuations is more gradual for faster rates of change of parameter, and the other measures (SKEW and $K$) are also able to predict the tipping before $\eta'_{rms}$ and VAR increases.

\begin{figure*}
    \centering
    \includegraphics[width=0.8\textwidth]{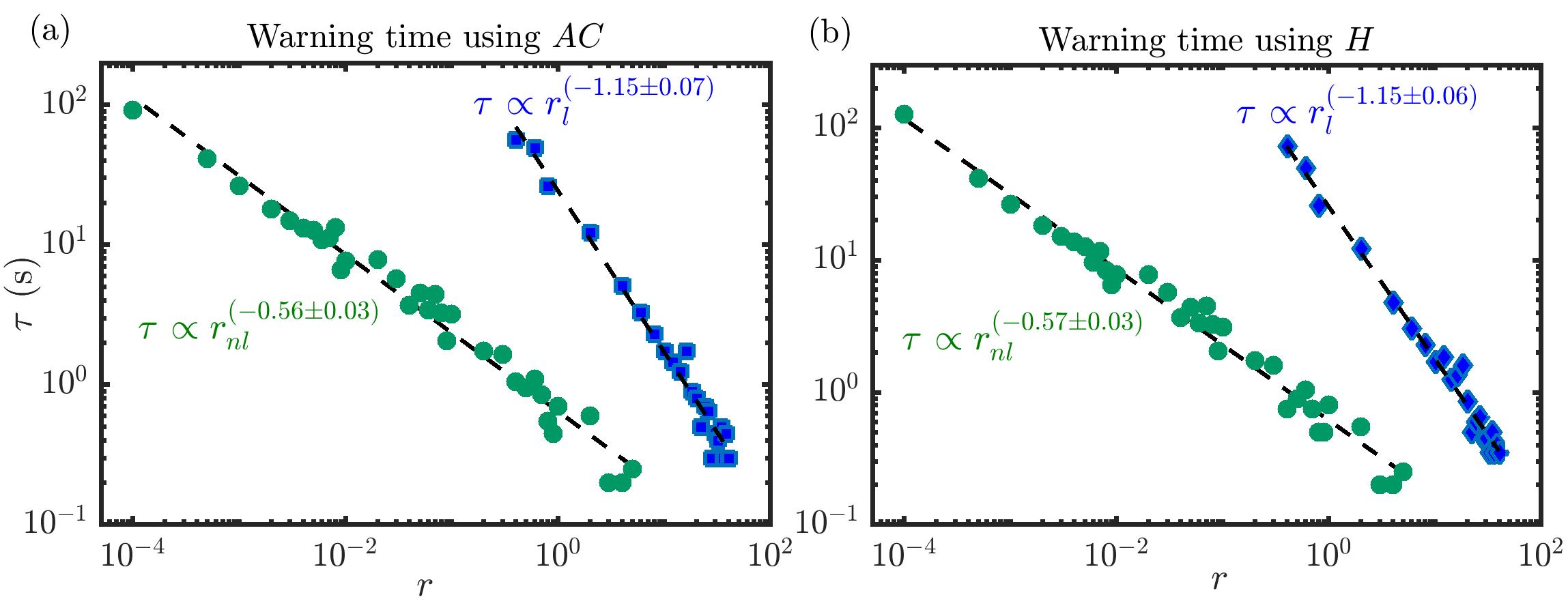}
    \caption{Warning time ($\tau$) obtained from (a) AC and (b) $H$ are plotted as a function of $r$ in a double logarithmic plot for both linear (blue colour) and quadratic (green colour) variation of $\alpha$. We observe a power law relation between $\tau$ and AC and $\tau$ and $H$. The threshold chosen for obtaining warning is 0.4 for AC and 0.1 for $H$.}
    \label{fig12}
\end{figure*}

Next, we analyze the warning time obtained using AC and $H$ for different rates of change of $\alpha$. The warning time reduces with the rate of change of parameter following an inverse power law relation. We show the scaling for linear and nonlinear variation of the parameter as $\alpha = \alpha_0 +r_l \ t$ and $\alpha = \alpha_0 +r_{nl} \ t^2$ (Fig.~\ref{fig12}). \textcolor{black}{The trends are qualitatively the same, even though} the obtained exponents are different from those of experiments. This difference in the exponent could be because of multiple reasons. The \textcolor{black}{transition} depends on the rate of change of parameter, the initial value of the parameter and the type and intensity of noise present in the system. Also, the other parameters such as the $\beta$, $\kappa$ and $\gamma$ can be varied to adjust the model to represent the experimental system better. In the current study, we vary $V$ linearly; ideally, $P$ has to vary as $P = V(t)^2/R$, where $R$ is the resistance. However, $R$ can change slightly with temperature as $V$ increases. 
In real systems, as we vary one parameter in the system, there can be multiple parameters changing simultaneously. Hence, the actual control parameter in practical systems can be a combination of multiple parameters. We do not attempt to obtain the exact
parameters and condition, but only aim for a qualitative match between the experiment and the model.

\subsection{Relation between lag-1 autocorrelation and variance close to the tipping point}

\begin{figure}[h!]
\centering
    \includegraphics[width=0.5\textwidth]{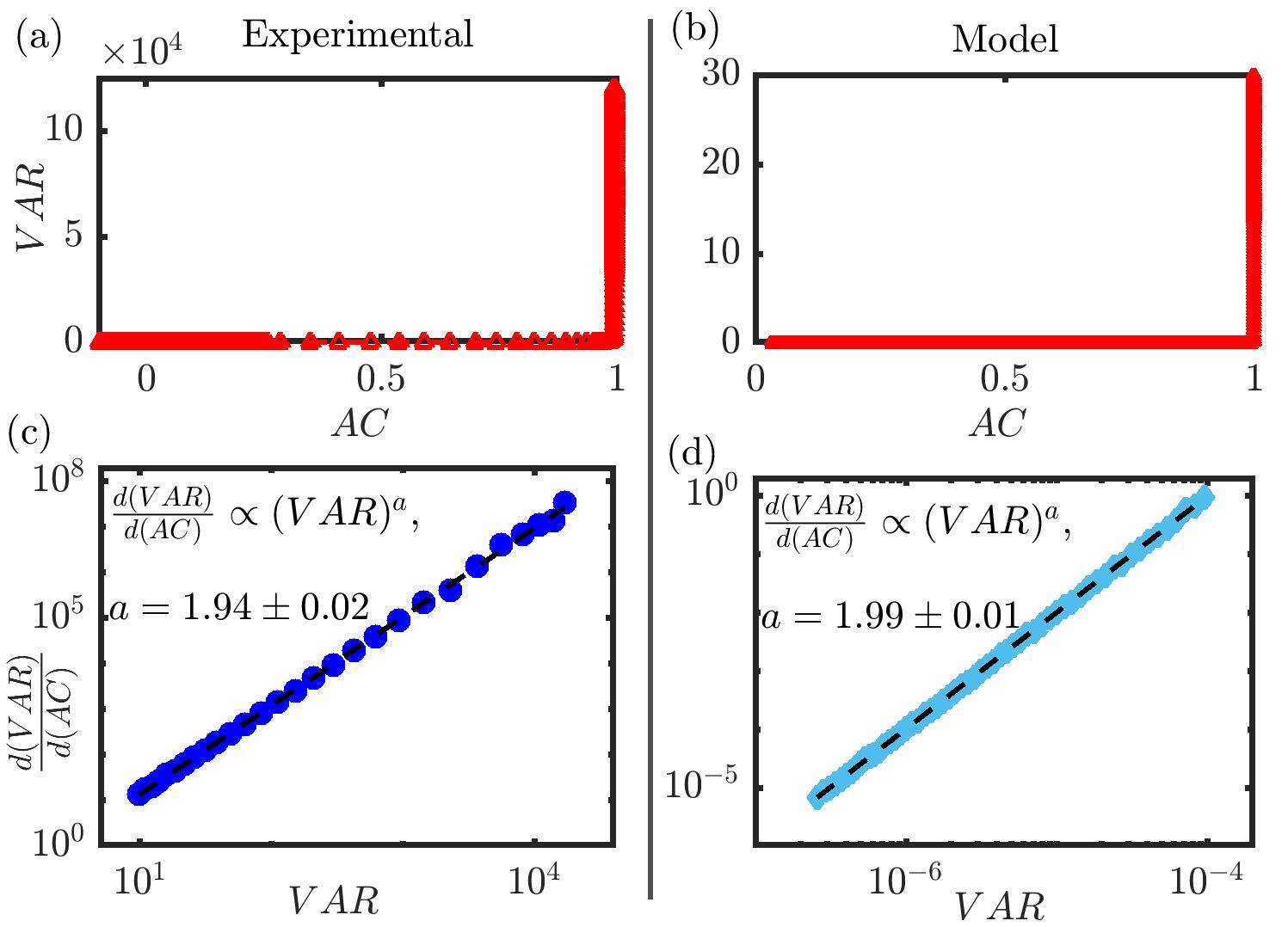}
    \caption{(a-b) Increase in the variance of fluctuations as a function of AC shows a curve increasing faster than exponential (hyperexponential). (c-d) To find the exponent of the hyperexponential function, a linear fit is done on the double logarithmic plot of $\frac{d(A)}{d(AC)}$ \textit{vs}. VAR.} 
    \label{fig13}
\end{figure}

Critical slowing down leads to an increased autocorrelation and variance of fluctuations in a stochastically perturbed system approaching a bifurcation \cite{scheffer2009early}. Near the critical point, AC tends to one, and VAR tends to infinity. Here, we examine the variation of lag-1 autocorrelation (AC) with the variance of fluctuations close to the tipping where the AC grows and saturates to 1. VAR stays nearly zero almost till the onset of TAI and then shoots up to really high values. At the same time, AC increases gradually during the transition. Here, we observe that AC scales with VAR following a hyperexponential relation \cite{varfolomeyev2001hyperexponential},
\begin{equation}
    \frac{d(VAR)}{d(AC)} = k(VAR)^a.
    \label{eq_hyper}
\end{equation} 
Refer to Fig.~\ref{fig13}a for the scaling obtained from experiments and Fig.~\ref{fig13}b for data from the model. 
The most common growth principle is exponential, where $a = 1$ in the above equation. To determine the exponent $a$, we fit $\frac{d(VAR)}{d(AC)}$ with $VAR$ in a double logarithmic plot (Figs.~\ref{fig13}c-d). The fitting is performed only for the data points in the middle of the curve in Fig.~\ref{fig13}a. The scaling shown in Fig.~\ref{fig13}c is a representative case of one experiment with a rate of 3 mV/s. The exponent is found to be $a$ = 1.94 $\pm$ 0.02 for the experiments and $a$ = 1.99 $\pm$ 0.01 for the model. 
On integrating Eq.~\ref{eq_hyper}, we get $VAR = [a_1(k\ AC+constant)]^{(1/a_1)}$, where $a_1 = -a +1$ and $a \sim 2$ from the Fig.~\ref{fig13}. Finally, we have empirically found a relation between AC and VAR for dynamic Hopf bifurcation as follows VAR $\propto \frac{-1}{k\ AC+constant}$. If we apply the limits approaching the tipping point: VAR tends to infinity as AC tends to one, we get the constant as $-k$ and the final expression as,
\begin{equation}
VAR \propto \frac{1}{1-AC}.
\end{equation}
We observe the same scaling exponent for linear and nonlinear variation of the control parameter in the model. The scaling relation seems to be independent of the functional form of the parameter. This hyperexponential scaling observed in experiments and model irrespective of the rate of change of parameter and the functional form of control parameter suggests that this can be a scaling during the occurrence of dynamic Hopf bifurcation.

\section{Conclusion}
We study several early warning signals for \textcolor{black}{critical transitions} in a thermoacoustic system. 
Compared to the quasi-static bifurcation, the onset of tipping is delayed \textcolor{black}{when the control parameter is varied continuously at a finite rate}. We confirm the observation of increased delay with increase in the rate of change of control parameter. By analyzing the performance of various early warning signals, we observe that the variance, kurtosis and skewness do not provide \textcolor{black}{adequate} warning; they change only when $p'_{rms}$ rises. 
The lag-1 autocorrelation and the Hurst exponent are able to predict \textcolor{black}{the transition} well-before the tipping point. We confirmed this observation by performing experiments at different rates of change of control parameter. For slower rates, AC and $H$ give more warning time compared to \textcolor{black}{faster} rates, even though we are relatively close to the transition in terms of the parameter. On the other hand, for faster rates where we have relatively lesser time to initiate control measures, AC and $H$ capture the tipping at a parameter value which is well ahead of the tipping point. Furthermore, we notice that the warning time reduces with the rate of change of parameter following an inverse power law relation. \textcolor{black}{Then, we perform a similar analysis for a} noisy Hopf bifurcation model. The qualitative features of the EWS using the lag-1 autocorrelation and the Hurst exponent are captured using the model.
Finally, we empirically obtained a relation between lag-1 autocorrelation and the variance of fluctuations for dynamic Hopf bifurcation. This hyperexponential scaling is found to be independent of the functional form of \textcolor{black}{variation of} the control parameter .

\begin{acknowledgments}
We acknowledge the discussions and help from Dr. Vishnu R. Unni, Dr. Gopalakrishnan and Dr. Samadhan Pawar. I. P. is grateful to the Ministry of Human Resource Development, India and Indian Institute of Technology Madras for providing research assistantship. R. I. S. acknowledges the funding from J.C. Bose fellowship (JCB/2018/000034/SSC).
\end{acknowledgments}

\section*{DATA AVAILABILITY}
The data that support the findings of this study are available from the corresponding author upon reasonable request.

\appendix
\section{Methodology of computation of early warning measures}
We provide an overview of the different EWS used in the present study. 
\subsection{Root mean square }
The root mean square ($rms$) of a time series, $p(t)$, is defined as the square root of the mean square or the arithmetic mean of the squares of all the elements in the time series. It is also known as the quadratic mean.
\begin{equation}
p_{\mathrm{rms}}=\sqrt{\frac{1}{N}\left(\sum_{i=1}^{N} p_i^{2}\right)}
\end{equation}
where \textit{N} is the total number of data points. 
Root mean square value has been widely used in many engineering applications as a first step to check the sudden increase in the amplitude of fluctuations about the mean.

\subsection{EWS based on Critical slowing down}
Systems approaching a transition where the current state becomes unstable and transitions to another state shows slow response to small external perturbations. This phenomenon of slow return rate is known as critical slowing down \cite{strogatz1994nonlinear} and can be detected by increased autocorrelation and variance of fluctuations \cite{scheffer2009early, dakos2012robustness}.
\subsubsection{Autocorrelation}
Autocorrelation is the correlation of a signal, $p(t)$, with a delayed copy of itself as a function of delay ($\tau$). It is defined as follows:
\begin{equation}
AC(\tau)=\frac{\sum_{i=1}^{N} p(i)\ {p(i-\tau)}}{{\sigma^2} }
\end{equation}
In the current study, we consider lag-1 autocorrelation (AC) which computes the correlation between values that are one time step apart. Critical slowing down leads to an increase in the short term memory of a system and can captured by correlation at low lags. 
\subsubsection{Variance}
Variance (VAR) is the expectation of the squared differences from its mean. VAR measures how the data is spread out from their average value and it is the second moment of the distribution.
\begin{equation}
\operatorname{VAR}=\frac{1}{N} \sum_{i=1}^{N}\left(p_{i}-\overline{p}\right)^{2}
\end{equation}
where $\overline{p}$ is the mean and \textit{N} is the number of data points. 

\subsection{Skewness and kurtosis}
Skewness (SKEW) is a measure of symmetry or the lack of symmetry of the probability distribution of data about its mean. A distribution is symmetric if it looks the same to the left and right of the mean value. The skewness for a normal distribution is zero and skewness will be near zero for any symmetric data. The skewness of a random variable \textit{X} is its third moment. 
\begin{equation}
SKEW = \frac{\sum_{i=1}^{N}(p_i- \overline{p})^3/N}{\sigma^3}   
\end{equation}
where $\overline{p}$ is the mean, $\sigma$ is the standard deviation, and \textit{N} is the number of data points. Negative SKEW indicates that the distribution is skewed left and positive SKEW indicates that the distribution is skewed right. Skewed left means that the left tail is longer than the right tail and vice versa. Close to a tipping point, probability distributions may become asymmetric with a non-zero skewness \cite{guttal2008changing}. 
Kurtosis ($K$) gives information about whether the probability distribution of the data has heavy tails, or is more centred compared to a normal distribution. The kurtosis of a normal distribution is 3, and higher kurtosis indicates more outliers or heavy tail in the data. If Kurtosis is less than 3, it means that the distribution has lesser outliers compared to the normal distribution.
The kurtosis is the fourth standardized moment and is defined as follows:
\begin{equation}
 K =\frac{\sum_{i=1}^{N}\left(p_{i}-\overline{p}\right)^{4} / N}{\sigma^{4}}
\end{equation}
where $\overline{p}$ is the mean, $\sigma$ is the standard deviation and \textit{N} is the number of data points. 

\begin{figure*}
\centering
    \includegraphics[width=0.75\textwidth]{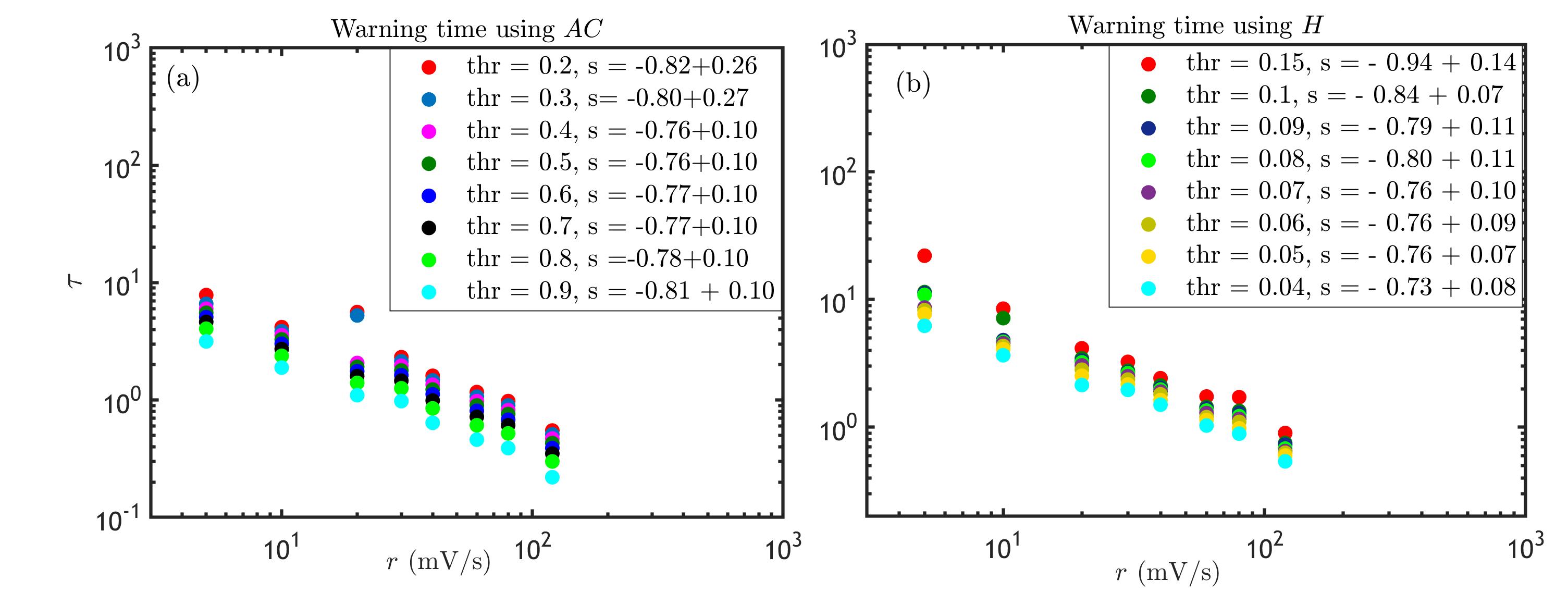}
    \caption{The inverse power law scaling between the warning time and rate of change of parameter is plotted for different values of thresholds for (a) AC and (b) $H$.} 
    \label{figA1}
\end{figure*}
\subsection{Hurst exponent}
The Hurst exponent ($H$) is related to the fractal dimension (\textit{D}) as $D=2-H$, in the case of a time series. Generally, \textit{H} has values between 0 and 1 for a time series of fractal dimension between 1 and 2. 
We calculate $H$ following the algorithm of Multifractal detrended fluctuation analysis \cite{kantelhardt2002multifractal, ihlen2012introduction}.

\subsubsection*{Multifractal detrended fluctuation analysis (MFDFA)}

First, we take a time series $x(t)$ of length \textit{N}. The mean subtracted cumulative deviate series $Y(k)$ can be defined as,
\begin{equation}
    Y(k)= \sum_{t=1}^{k} [x_t- \bar{x}], \qquad k = 1, 2,..., N,
\end{equation}
where $\bar{x}$ is the mean of the time series. We divide the deviate series $Y(k)$ into $N_w = [N/w]$ non-overlapping segments of equal length $w$, where $[N/w]$ represents the greatest integer function. 
Then, we find a polynomial fit ($\bar{Y_i}$) for each of these segments \textit{i} and we subtract the polynomial fit from the deviate series ($Y_i$) to obtain the fluctuations. In the present study, we use a polynomial fit of order 1. Then, the variance of fluctuations is determined as,
\begin{equation}
    F^2(w,i) = \frac{1}{w}\Big[ \sum_{t=1}^{w}(Y_i(t)-\bar{Y_i})^2 \Big ], 
\end{equation}
for each segment $i= 1, 2,...,N_w$.

\noindent The structure function of order 2 and span $w$, $F_w^2$ is defined as follows:
\begin{equation}
    F_w^2 = \Big[\frac{1}{N_w} \sum_{i=1}^{N_w}F^2(w,i) \Big ]^{1/2}.
\end{equation}

We repeat the same steps for different time scales or span $w$ and plot the variation of $F_w^2$ with the span $w$ in a logarithmic scale. The slope of the linear regime for a range of span sizes $w$ is known as the Hurst exponent (\textit{H}).

\section{Robustness of EWS with the threshold}
We observe that the inverse power law scaling between the warning time and rate of change of parameter is consistent with almost the same exponents for different values of thresholds of EWS as shown in Fig.~\ref{figA1}.
\begin{figure}[h!]
\centering
    \includegraphics[width=0.52\textwidth]{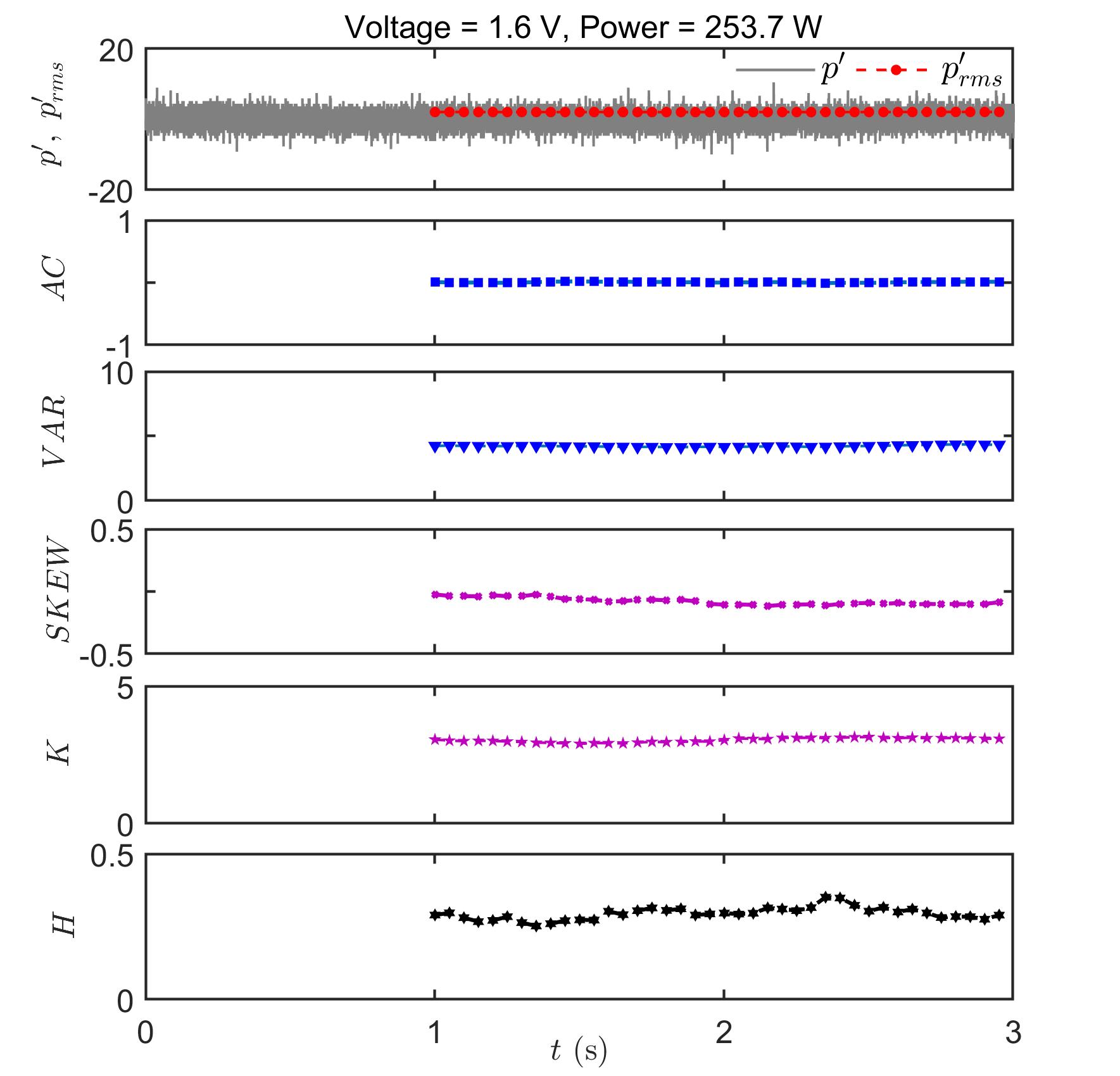}
    \caption{Acoustic pressure fluctuations ($p'$) acquired for a constant value of control parameter (voltage = 1.6 V and heater power = 253.7 W), lower than the Hopf point. The time series contains only low amplitude aperiodic fluctuations, and there is no transition to limit cycle oscillations. The \textit{rms} of pressure fluctuations is plotted along with the signal in the top plot. The corresponding variation of all the EWS is shown as subplots. We calculate the EWS for a moving window of 1 s with an overlap of 0.98 s. $p’_{rms}$, AC, VAR, SKEW, $K$ and $H$ stay nearly constant for the total duration of the experiment, and we do not observe any significant change in the values of EWS.} 
    \label{figA2}
\end{figure}
\section{Analysis to check false warnings}
To check for false warnings, we calculate EWS for data for cases where transition to thermoacoustic instability does not occur. Here, we have the time series data acquired for constant values of voltage (quasi-static experiments). Figure \ref{figA2} shows one such data and the corresponding variation of all the EWS. This is a representative case of data for voltage = 1.6 V and heater power = 253.7 W, and we have confirmed these observations by analysing many data sets which are taken for quasi-static experiments.  We wait at a particular value of control parameter far below the Hopf point, and there is no transition to limit cycle oscillations as expected. All the EWS calculated for a moving window show constant values indicative of a low amplitude aperiodic state, for the entire length of the data. Hence, we do not observe any false warnings for these EWS.

\bibliography{aipsamp}

\providecommand{\noopsort}[1]{}\providecommand{\singleletter}[1]{#1}%
\begin{thebibliography}{65}%
\makeatletter
\providecommand \@ifxundefined [1]{%
 \@ifx{#1\undefined}
}%
\providecommand \@ifnum [1]{%
 \ifnum #1\expandafter \@firstoftwo
 \else \expandafter \@secondoftwo
 \fi
}%
\providecommand \@ifx [1]{%
 \ifx #1\expandafter \@firstoftwo
 \else \expandafter \@secondoftwo
 \fi
}%
\providecommand \natexlab [1]{#1}%
\providecommand \enquote  [1]{``#1''}%
\providecommand \bibnamefont  [1]{#1}%
\providecommand \bibfnamefont [1]{#1}%
\providecommand \citenamefont [1]{#1}%
\providecommand \href@noop [0]{\@secondoftwo}%
\providecommand \href [0]{\begingroup \@sanitize@url \@href}%
\providecommand \@href[1]{\@@startlink{#1}\@@href}%
\providecommand \@@href[1]{\endgroup#1\@@endlink}%
\providecommand \@sanitize@url [0]{\catcode `\\12\catcode `\$12\catcode
  `\&12\catcode `\#12\catcode `\^12\catcode `\_12\catcode `\%12\relax}%
\providecommand \@@startlink[1]{}%
\providecommand \@@endlink[0]{}%
\providecommand \url  [0]{\begingroup\@sanitize@url \@url }%
\providecommand \@url [1]{\endgroup\@href {#1}{\urlprefix }}%
\providecommand \urlprefix  [0]{URL }%
\providecommand \Eprint [0]{\href }%
\providecommand \doibase [0]{http://dx.doi.org/}%
\providecommand \selectlanguage [0]{\@gobble}%
\providecommand \bibinfo  [0]{\@secondoftwo}%
\providecommand \bibfield  [0]{\@secondoftwo}%
\providecommand \translation [1]{[#1]}%
\providecommand \BibitemOpen [0]{}%
\providecommand \bibitemStop [0]{}%
\providecommand \bibitemNoStop [0]{.\EOS\space}%
\providecommand \EOS [0]{\spacefactor3000\relax}%
\providecommand \BibitemShut  [1]{\csname bibitem#1\endcsname}%
\let\auto@bib@innerbib\@empty
\bibitem [{\citenamefont {Lenton}\ \emph {et~al.}(2008)\citenamefont {Lenton},
  \citenamefont {Held}, \citenamefont {Kriegler}, \citenamefont {Hall},
  \citenamefont {Lucht}, \citenamefont {Rahmstorf},\ and\ \citenamefont
  {Schellnhuber}}]{lenton2008tipping}%
  \BibitemOpen
  \bibfield  {author} {\bibinfo {author} {\bibfnamefont {T.~M.}\ \bibnamefont
  {Lenton}}, \bibinfo {author} {\bibfnamefont {H.}~\bibnamefont {Held}},
  \bibinfo {author} {\bibfnamefont {E.}~\bibnamefont {Kriegler}}, \bibinfo
  {author} {\bibfnamefont {J.~W.}\ \bibnamefont {Hall}}, \bibinfo {author}
  {\bibfnamefont {W.}~\bibnamefont {Lucht}}, \bibinfo {author} {\bibfnamefont
  {S.}~\bibnamefont {Rahmstorf}}, \ and\ \bibinfo {author} {\bibfnamefont
  {H.~J.}\ \bibnamefont {Schellnhuber}},\ }\bibfield  {title} {\enquote
  {\bibinfo {title} {Tipping elements in the earth's climate system},}\
  }\href@noop {} {\bibfield  {journal} {\bibinfo  {journal} {Proceedings of the
  national Academy of Sciences}\ }\textbf {\bibinfo {volume} {105}},\ \bibinfo
  {pages} {1786--1793} (\bibinfo {year} {2008})}\BibitemShut {NoStop}%
\bibitem [{\citenamefont {Scheffer}\ \emph {et~al.}(2001)\citenamefont
  {Scheffer}, \citenamefont {Carpenter}, \citenamefont {Foley}, \citenamefont
  {Folke},\ and\ \citenamefont {Walker}}]{scheffer2001catastrophic}%
  \BibitemOpen
  \bibfield  {author} {\bibinfo {author} {\bibfnamefont {M.}~\bibnamefont
  {Scheffer}}, \bibinfo {author} {\bibfnamefont {S.}~\bibnamefont {Carpenter}},
  \bibinfo {author} {\bibfnamefont {J.~A.}\ \bibnamefont {Foley}}, \bibinfo
  {author} {\bibfnamefont {C.}~\bibnamefont {Folke}}, \ and\ \bibinfo {author}
  {\bibfnamefont {B.}~\bibnamefont {Walker}},\ }\bibfield  {title} {\enquote
  {\bibinfo {title} {Catastrophic shifts in ecosystems},}\ }\href@noop {}
  {\bibfield  {journal} {\bibinfo  {journal} {Nature}\ }\textbf {\bibinfo
  {volume} {413}},\ \bibinfo {pages} {591} (\bibinfo {year}
  {2001})}\BibitemShut {NoStop}%
\bibitem [{\citenamefont {Sornette}\ and\ \citenamefont
  {Johansen}(1997)}]{sornette1997large}%
  \BibitemOpen
  \bibfield  {author} {\bibinfo {author} {\bibfnamefont {D.}~\bibnamefont
  {Sornette}}\ and\ \bibinfo {author} {\bibfnamefont {A.}~\bibnamefont
  {Johansen}},\ }\bibfield  {title} {\enquote {\bibinfo {title} {Large
  financial crashes},}\ }\href@noop {} {\bibfield  {journal} {\bibinfo
  {journal} {Physica A: Statistical Mechanics and its Applications}\ }\textbf
  {\bibinfo {volume} {245}},\ \bibinfo {pages} {411--422} (\bibinfo {year}
  {1997})}\BibitemShut {NoStop}%
\bibitem [{\citenamefont {Venegas}\ \emph {et~al.}(2005)\citenamefont
  {Venegas}, \citenamefont {Winkler}, \citenamefont {Musch}, \citenamefont
  {Melo}, \citenamefont {Layfield}, \citenamefont {Tgavalekos}, \citenamefont
  {Fischman}, \citenamefont {Callahan}, \citenamefont {Bellani},\ and\
  \citenamefont {Harris}}]{venegas2005self}%
  \BibitemOpen
  \bibfield  {author} {\bibinfo {author} {\bibfnamefont {J.~G.}\ \bibnamefont
  {Venegas}}, \bibinfo {author} {\bibfnamefont {T.}~\bibnamefont {Winkler}},
  \bibinfo {author} {\bibfnamefont {G.}~\bibnamefont {Musch}}, \bibinfo
  {author} {\bibfnamefont {M.~F.~V.}\ \bibnamefont {Melo}}, \bibinfo {author}
  {\bibfnamefont {D.}~\bibnamefont {Layfield}}, \bibinfo {author}
  {\bibfnamefont {N.}~\bibnamefont {Tgavalekos}}, \bibinfo {author}
  {\bibfnamefont {A.~J.}\ \bibnamefont {Fischman}}, \bibinfo {author}
  {\bibfnamefont {R.~J.}\ \bibnamefont {Callahan}}, \bibinfo {author}
  {\bibfnamefont {G.}~\bibnamefont {Bellani}}, \ and\ \bibinfo {author}
  {\bibfnamefont {R.~S.}\ \bibnamefont {Harris}},\ }\bibfield  {title}
  {\enquote {\bibinfo {title} {Self-organized patchiness in asthma as a prelude
  to catastrophic shifts},}\ }\href@noop {} {\bibfield  {journal} {\bibinfo
  {journal} {Nature}\ }\textbf {\bibinfo {volume} {434}},\ \bibinfo {pages}
  {777} (\bibinfo {year} {2005})}\BibitemShut {NoStop}%
\bibitem [{\citenamefont {Litt}\ \emph {et~al.}(2001)\citenamefont {Litt},
  \citenamefont {Esteller}, \citenamefont {Echauz}, \citenamefont
  {D'Alessandro}, \citenamefont {Shor}, \citenamefont {Henry}, \citenamefont
  {Pennell}, \citenamefont {Epstein}, \citenamefont {Bakay}, \citenamefont
  {Dichter} \emph {et~al.}}]{litt2001epileptic}%
  \BibitemOpen
  \bibfield  {author} {\bibinfo {author} {\bibfnamefont {B.}~\bibnamefont
  {Litt}}, \bibinfo {author} {\bibfnamefont {R.}~\bibnamefont {Esteller}},
  \bibinfo {author} {\bibfnamefont {J.}~\bibnamefont {Echauz}}, \bibinfo
  {author} {\bibfnamefont {M.}~\bibnamefont {D'Alessandro}}, \bibinfo {author}
  {\bibfnamefont {R.}~\bibnamefont {Shor}}, \bibinfo {author} {\bibfnamefont
  {T.}~\bibnamefont {Henry}}, \bibinfo {author} {\bibfnamefont
  {P.}~\bibnamefont {Pennell}}, \bibinfo {author} {\bibfnamefont
  {C.}~\bibnamefont {Epstein}}, \bibinfo {author} {\bibfnamefont
  {R.}~\bibnamefont {Bakay}}, \bibinfo {author} {\bibfnamefont
  {M.}~\bibnamefont {Dichter}},  \emph {et~al.},\ }\bibfield  {title} {\enquote
  {\bibinfo {title} {Epileptic seizures may begin hours in advance of clinical
  onset: a report of five patients},}\ }\href@noop {} {\bibfield  {journal}
  {\bibinfo  {journal} {Neuron}\ }\textbf {\bibinfo {volume} {30}},\ \bibinfo
  {pages} {51--64} (\bibinfo {year} {2001})}\BibitemShut {NoStop}%
\bibitem [{\citenamefont {McSharry}, \citenamefont {Smith},\ and\ \citenamefont
  {Tarassenko}(2003)}]{mcsharry2003prediction}%
  \BibitemOpen
  \bibfield  {author} {\bibinfo {author} {\bibfnamefont {P.~E.}\ \bibnamefont
  {McSharry}}, \bibinfo {author} {\bibfnamefont {L.~A.}\ \bibnamefont {Smith}},
  \ and\ \bibinfo {author} {\bibfnamefont {L.}~\bibnamefont {Tarassenko}},\
  }\bibfield  {title} {\enquote {\bibinfo {title} {Prediction of epileptic
  seizures: are nonlinear methods relevant?}}\ }\href@noop {} {\bibfield
  {journal} {\bibinfo  {journal} {Nature medicine}\ }\textbf {\bibinfo {volume}
  {9}},\ \bibinfo {pages} {241} (\bibinfo {year} {2003})}\BibitemShut {NoStop}%
\bibitem [{\citenamefont {Carpenter}, \citenamefont {Ludwig},\ and\
  \citenamefont {Brock}(1999)}]{carpenter1999management}%
  \BibitemOpen
  \bibfield  {author} {\bibinfo {author} {\bibfnamefont {S.~R.}\ \bibnamefont
  {Carpenter}}, \bibinfo {author} {\bibfnamefont {D.}~\bibnamefont {Ludwig}}, \
  and\ \bibinfo {author} {\bibfnamefont {W.~A.}\ \bibnamefont {Brock}},\
  }\bibfield  {title} {\enquote {\bibinfo {title} {Management of eutrophication
  for lakes subject to potentially irreversible change},}\ }\href@noop {}
  {\bibfield  {journal} {\bibinfo  {journal} {Ecological applications}\
  }\textbf {\bibinfo {volume} {9}},\ \bibinfo {pages} {751--771} (\bibinfo
  {year} {1999})}\BibitemShut {NoStop}%
\bibitem [{\citenamefont {National-Research-Councils}\ and\ \citenamefont
  {other}(2007)}]{national2007new}%
  \BibitemOpen
  \bibfield  {author} {\bibinfo {author} {\bibnamefont
  {National-Research-Councils}}\ and\ \bibinfo {author} {\bibnamefont
  {other}},\ }\href@noop {} {\emph {\bibinfo {title} {New directions for
  understanding systemic risk: a report on a conference cosponsored by the
  Federal Reserve Bank of New York and the National Academy of Sciences}}}\
  (\bibinfo  {publisher} {National Academies Press},\ \bibinfo {year}
  {2007})\BibitemShut {NoStop}%
\bibitem [{\citenamefont {May}, \citenamefont {Levin},\ and\ \citenamefont
  {Sugihara}(2008)}]{may2008ecology}%
  \BibitemOpen
  \bibfield  {author} {\bibinfo {author} {\bibfnamefont {R.~M.}\ \bibnamefont
  {May}}, \bibinfo {author} {\bibfnamefont {S.~A.}\ \bibnamefont {Levin}}, \
  and\ \bibinfo {author} {\bibfnamefont {G.}~\bibnamefont {Sugihara}},\
  }\bibfield  {title} {\enquote {\bibinfo {title} {Ecology for bankers},}\
  }\href@noop {} {\bibfield  {journal} {\bibinfo  {journal} {Nature}\ }\textbf
  {\bibinfo {volume} {451}},\ \bibinfo {pages} {893--894} (\bibinfo {year}
  {2008})}\BibitemShut {NoStop}%
\bibitem [{\citenamefont {Ashwin}\ \emph {et~al.}(2012)\citenamefont {Ashwin},
  \citenamefont {Wieczorek}, \citenamefont {Vitolo},\ and\ \citenamefont
  {Cox}}]{ashwin2012tipping}%
  \BibitemOpen
  \bibfield  {author} {\bibinfo {author} {\bibfnamefont {P.}~\bibnamefont
  {Ashwin}}, \bibinfo {author} {\bibfnamefont {S.}~\bibnamefont {Wieczorek}},
  \bibinfo {author} {\bibfnamefont {R.}~\bibnamefont {Vitolo}}, \ and\ \bibinfo
  {author} {\bibfnamefont {P.}~\bibnamefont {Cox}},\ }\bibfield  {title}
  {\enquote {\bibinfo {title} {Tipping points in open systems: bifurcation,
  noise-induced and rate-dependent examples in the climate system},}\
  }\href@noop {} {\bibfield  {journal} {\bibinfo  {journal} {Philosophical
  Transactions of the Royal Society A: Mathematical, Physical and Engineering
  Sciences}\ }\textbf {\bibinfo {volume} {370}},\ \bibinfo {pages} {1166--1184}
  (\bibinfo {year} {2012})}\BibitemShut {NoStop}%
\bibitem [{\citenamefont {Tony}\ \emph {et~al.}(2017)\citenamefont {Tony},
  \citenamefont {Subarna}, \citenamefont {Syamkumar}, \citenamefont {Sudha},
  \citenamefont {Akshay}, \citenamefont {Gopalakrishnan}, \citenamefont
  {Surovyatkina},\ and\ \citenamefont {Sujith}}]{tony2017experimental}%
  \BibitemOpen
  \bibfield  {author} {\bibinfo {author} {\bibfnamefont {J.}~\bibnamefont
  {Tony}}, \bibinfo {author} {\bibfnamefont {S.}~\bibnamefont {Subarna}},
  \bibinfo {author} {\bibfnamefont {K.~S.}\ \bibnamefont {Syamkumar}}, \bibinfo
  {author} {\bibfnamefont {G.}~\bibnamefont {Sudha}}, \bibinfo {author}
  {\bibfnamefont {S.}~\bibnamefont {Akshay}}, \bibinfo {author} {\bibfnamefont
  {E.~A.}\ \bibnamefont {Gopalakrishnan}}, \bibinfo {author} {\bibfnamefont
  {E.}~\bibnamefont {Surovyatkina}}, \ and\ \bibinfo {author} {\bibfnamefont
  {R.~I.}\ \bibnamefont {Sujith}},\ }\bibfield  {title} {\enquote {\bibinfo
  {title} {Experimental investigation on preconditioned rate induced tipping in
  a thermoacoustic system},}\ }\href@noop {} {\bibfield  {journal} {\bibinfo
  {journal} {Scientific reports}\ }\textbf {\bibinfo {volume} {7}},\ \bibinfo
  {pages} {1--7} (\bibinfo {year} {2017})}\BibitemShut {NoStop}%
\bibitem [{\citenamefont {Baer}, \citenamefont {Erneux},\ and\ \citenamefont
  {Rinzel}(1989)}]{baer1989slow}%
  \BibitemOpen
  \bibfield  {author} {\bibinfo {author} {\bibfnamefont {S.~M.}\ \bibnamefont
  {Baer}}, \bibinfo {author} {\bibfnamefont {T.}~\bibnamefont {Erneux}}, \ and\
  \bibinfo {author} {\bibfnamefont {J.}~\bibnamefont {Rinzel}},\ }\bibfield
  {title} {\enquote {\bibinfo {title} {The slow passage through a hopf
  bifurcation: delay, memory effects, and resonance},}\ }\href@noop {}
  {\bibfield  {journal} {\bibinfo  {journal} {SIAM Journal on Applied
  mathematics}\ }\textbf {\bibinfo {volume} {49}},\ \bibinfo {pages} {55--71}
  (\bibinfo {year} {1989})}\BibitemShut {NoStop}%
\bibitem [{\citenamefont {Bonciolini}\ \emph {et~al.}(2018)\citenamefont
  {Bonciolini}, \citenamefont {Ebi}, \citenamefont {Boujo},\ and\ \citenamefont
  {Noiray}}]{bonciolini2018experiments}%
  \BibitemOpen
  \bibfield  {author} {\bibinfo {author} {\bibfnamefont {G.}~\bibnamefont
  {Bonciolini}}, \bibinfo {author} {\bibfnamefont {D.}~\bibnamefont {Ebi}},
  \bibinfo {author} {\bibfnamefont {E.}~\bibnamefont {Boujo}}, \ and\ \bibinfo
  {author} {\bibfnamefont {N.}~\bibnamefont {Noiray}},\ }\bibfield  {title}
  {\enquote {\bibinfo {title} {Experiments and modelling of rate-dependent
  transition delay in a stochastic subcritical bifurcation},}\ }\href@noop {}
  {\bibfield  {journal} {\bibinfo  {journal} {Royal Society open science}\
  }\textbf {\bibinfo {volume} {5}},\ \bibinfo {pages} {172078} (\bibinfo {year}
  {2018})}\BibitemShut {NoStop}%
\bibitem [{\citenamefont {Park}, \citenamefont {Do},\ and\ \citenamefont
  {Lopez}(2011)}]{park2011slow}%
  \BibitemOpen
  \bibfield  {author} {\bibinfo {author} {\bibfnamefont {Y.}~\bibnamefont
  {Park}}, \bibinfo {author} {\bibfnamefont {Y.}~\bibnamefont {Do}}, \ and\
  \bibinfo {author} {\bibfnamefont {J.~M.}\ \bibnamefont {Lopez}},\ }\bibfield
  {title} {\enquote {\bibinfo {title} {Slow passage through resonance},}\
  }\href@noop {} {\bibfield  {journal} {\bibinfo  {journal} {Physical Review
  E}\ }\textbf {\bibinfo {volume} {84}},\ \bibinfo {pages} {056604} (\bibinfo
  {year} {2011})}\BibitemShut {NoStop}%
\bibitem [{\citenamefont {Berglund}(2000)}]{berglund2000dynamic}%
  \BibitemOpen
  \bibfield  {author} {\bibinfo {author} {\bibfnamefont {N.}~\bibnamefont
  {Berglund}},\ }\bibfield  {title} {\enquote {\bibinfo {title} {Dynamic
  bifurcations: Hysteresis, scaling laws and feedback control},}\ }\href@noop
  {} {\bibfield  {journal} {\bibinfo  {journal} {Progress of Theoretical
  Physics Supplement}\ }\textbf {\bibinfo {volume} {139}},\ \bibinfo {pages}
  {325--336} (\bibinfo {year} {2000})}\BibitemShut {NoStop}%
\bibitem [{\citenamefont {Majumdar}\ \emph {et~al.}(2013)\citenamefont
  {Majumdar}, \citenamefont {Ockendon}, \citenamefont {Howell},\ and\
  \citenamefont {Surovyatkina}}]{majumdar2013transitions}%
  \BibitemOpen
  \bibfield  {author} {\bibinfo {author} {\bibfnamefont {A.}~\bibnamefont
  {Majumdar}}, \bibinfo {author} {\bibfnamefont {J.}~\bibnamefont {Ockendon}},
  \bibinfo {author} {\bibfnamefont {P.}~\bibnamefont {Howell}}, \ and\ \bibinfo
  {author} {\bibfnamefont {E.}~\bibnamefont {Surovyatkina}},\ }\bibfield
  {title} {\enquote {\bibinfo {title} {Transitions through critical
  temperatures in nematic liquid crystals},}\ }\href@noop {} {\bibfield
  {journal} {\bibinfo  {journal} {Physical Review E}\ }\textbf {\bibinfo
  {volume} {88}},\ \bibinfo {pages} {022501} (\bibinfo {year}
  {2013})}\BibitemShut {NoStop}%
\bibitem [{\citenamefont {Unni}\ \emph {et~al.}(2019)\citenamefont {Unni},
  \citenamefont {Gopalakrishnan}, \citenamefont {Syamkumar}, \citenamefont
  {Sujith}, \citenamefont {Surovyatkina},\ and\ \citenamefont
  {Kurths}}]{unni2019interplay}%
  \BibitemOpen
  \bibfield  {author} {\bibinfo {author} {\bibfnamefont {V.~R.}\ \bibnamefont
  {Unni}}, \bibinfo {author} {\bibfnamefont {E.~A.}\ \bibnamefont
  {Gopalakrishnan}}, \bibinfo {author} {\bibfnamefont {K.~S.}\ \bibnamefont
  {Syamkumar}}, \bibinfo {author} {\bibfnamefont {R.~I.}\ \bibnamefont
  {Sujith}}, \bibinfo {author} {\bibfnamefont {E.}~\bibnamefont
  {Surovyatkina}}, \ and\ \bibinfo {author} {\bibfnamefont {J.}~\bibnamefont
  {Kurths}},\ }\bibfield  {title} {\enquote {\bibinfo {title} {Interplay
  between random fluctuations and rate dependent phenomena at slow passage to
  limit-cycle oscillations in a bistable thermoacoustic system},}\ }\href@noop
  {} {\bibfield  {journal} {\bibinfo  {journal} {Chaos: An Interdisciplinary
  Journal of Nonlinear Science}\ }\textbf {\bibinfo {volume} {29}},\ \bibinfo
  {pages} {031102} (\bibinfo {year} {2019})}\BibitemShut {NoStop}%
\bibitem [{\citenamefont {Schroeder}(2009)}]{schroeder2009fractals}%
  \BibitemOpen
  \bibfield  {author} {\bibinfo {author} {\bibfnamefont {M.}~\bibnamefont
  {Schroeder}},\ }\href@noop {} {\emph {\bibinfo {title} {Fractals, chaos,
  power laws: Minutes from an infinite paradise}}}\ (\bibinfo  {publisher}
  {Courier Corporation},\ \bibinfo {year} {2009})\BibitemShut {NoStop}%
\bibitem [{\citenamefont {Scheffer}\ \emph {et~al.}(2009)\citenamefont
  {Scheffer}, \citenamefont {Bascompte}, \citenamefont {Brock}, \citenamefont
  {Brovkin}, \citenamefont {Carpenter}, \citenamefont {Dakos}, \citenamefont
  {Held}, \citenamefont {Van~Nes}, \citenamefont {Rietkerk},\ and\
  \citenamefont {Sugihara}}]{scheffer2009early}%
  \BibitemOpen
  \bibfield  {author} {\bibinfo {author} {\bibfnamefont {M.}~\bibnamefont
  {Scheffer}}, \bibinfo {author} {\bibfnamefont {J.}~\bibnamefont {Bascompte}},
  \bibinfo {author} {\bibfnamefont {W.~A.}\ \bibnamefont {Brock}}, \bibinfo
  {author} {\bibfnamefont {V.}~\bibnamefont {Brovkin}}, \bibinfo {author}
  {\bibfnamefont {S.~R.}\ \bibnamefont {Carpenter}}, \bibinfo {author}
  {\bibfnamefont {V.}~\bibnamefont {Dakos}}, \bibinfo {author} {\bibfnamefont
  {H.}~\bibnamefont {Held}}, \bibinfo {author} {\bibfnamefont {E.~H.}\
  \bibnamefont {Van~Nes}}, \bibinfo {author} {\bibfnamefont {M.}~\bibnamefont
  {Rietkerk}}, \ and\ \bibinfo {author} {\bibfnamefont {G.}~\bibnamefont
  {Sugihara}},\ }\bibfield  {title} {\enquote {\bibinfo {title} {Early-warning
  signals for critical transitions},}\ }\href@noop {} {\bibfield  {journal}
  {\bibinfo  {journal} {Nature}\ }\textbf {\bibinfo {volume} {461}},\ \bibinfo
  {pages} {53} (\bibinfo {year} {2009})}\BibitemShut {NoStop}%
\bibitem [{\citenamefont {Dakos}\ \emph {et~al.}(2008)\citenamefont {Dakos},
  \citenamefont {Scheffer}, \citenamefont {van Nes}, \citenamefont {Brovkin},
  \citenamefont {Petoukhov},\ and\ \citenamefont {Held}}]{dakos2008slowing}%
  \BibitemOpen
  \bibfield  {author} {\bibinfo {author} {\bibfnamefont {V.}~\bibnamefont
  {Dakos}}, \bibinfo {author} {\bibfnamefont {M.}~\bibnamefont {Scheffer}},
  \bibinfo {author} {\bibfnamefont {E.~H.}\ \bibnamefont {van Nes}}, \bibinfo
  {author} {\bibfnamefont {V.}~\bibnamefont {Brovkin}}, \bibinfo {author}
  {\bibfnamefont {V.}~\bibnamefont {Petoukhov}}, \ and\ \bibinfo {author}
  {\bibfnamefont {H.}~\bibnamefont {Held}},\ }\bibfield  {title} {\enquote
  {\bibinfo {title} {Slowing down as an early warning signal for abrupt climate
  change},}\ }\href@noop {} {\bibfield  {journal} {\bibinfo  {journal}
  {Proceedings of the National Academy of Sciences}\ }\textbf {\bibinfo
  {volume} {105}},\ \bibinfo {pages} {14308--14312} (\bibinfo {year}
  {2008})}\BibitemShut {NoStop}%
\bibitem [{\citenamefont {Wilkat}, \citenamefont {Rings},\ and\ \citenamefont
  {Lehnertz}(2019)}]{wilkat2019no}%
  \BibitemOpen
  \bibfield  {author} {\bibinfo {author} {\bibfnamefont {T.}~\bibnamefont
  {Wilkat}}, \bibinfo {author} {\bibfnamefont {T.}~\bibnamefont {Rings}}, \
  and\ \bibinfo {author} {\bibfnamefont {K.}~\bibnamefont {Lehnertz}},\
  }\bibfield  {title} {\enquote {\bibinfo {title} {No evidence for critical
  slowing down prior to human epileptic seizures},}\ }\href@noop {} {\bibfield
  {journal} {\bibinfo  {journal} {Chaos: An Interdisciplinary Journal of
  Nonlinear Science}\ }\textbf {\bibinfo {volume} {29}},\ \bibinfo {pages}
  {091104} (\bibinfo {year} {2019})}\BibitemShut {NoStop}%
\bibitem [{\citenamefont {Tsotsis}, \citenamefont {Sane},\ and\ \citenamefont
  {Lindstrom}(1988)}]{tsotsis1988bifurcation}%
  \BibitemOpen
  \bibfield  {author} {\bibinfo {author} {\bibfnamefont {T.~T.}\ \bibnamefont
  {Tsotsis}}, \bibinfo {author} {\bibfnamefont {R.~C.}\ \bibnamefont {Sane}}, \
  and\ \bibinfo {author} {\bibfnamefont {T.~H.}\ \bibnamefont {Lindstrom}},\
  }\bibfield  {title} {\enquote {\bibinfo {title} {Bifurcation behavior of a
  catalytic reaction due to a slowly varying parameter},}\ }\href@noop {}
  {\bibfield  {journal} {\bibinfo  {journal} {AIChE journal}\ }\textbf
  {\bibinfo {volume} {34}},\ \bibinfo {pages} {383--388} (\bibinfo {year}
  {1988})}\BibitemShut {NoStop}%
\bibitem [{\citenamefont {Kapila}(1981)}]{kapila1981arrhenius}%
  \BibitemOpen
  \bibfield  {author} {\bibinfo {author} {\bibfnamefont {A.~K.}\ \bibnamefont
  {Kapila}},\ }\bibfield  {title} {\enquote {\bibinfo {title} {Arrhenius
  systems: dynamics of jump due to slow passage through criticality},}\
  }\href@noop {} {\bibfield  {journal} {\bibinfo  {journal} {SIAM Journal on
  Applied Mathematics}\ }\textbf {\bibinfo {volume} {41}},\ \bibinfo {pages}
  {29--42} (\bibinfo {year} {1981})}\BibitemShut {NoStop}%
\bibitem [{\citenamefont {Bilinsky}\ and\ \citenamefont
  {Baer}(2018)}]{bilinsky2018slow}%
  \BibitemOpen
  \bibfield  {author} {\bibinfo {author} {\bibfnamefont {L.~M.}\ \bibnamefont
  {Bilinsky}}\ and\ \bibinfo {author} {\bibfnamefont {S.~M.}\ \bibnamefont
  {Baer}},\ }\bibfield  {title} {\enquote {\bibinfo {title} {Slow passage
  through a hopf bifurcation in excitable nerve cables: Spatial delays and
  spatial memory effects},}\ }\href@noop {} {\bibfield  {journal} {\bibinfo
  {journal} {Bulletin of mathematical biology}\ }\textbf {\bibinfo {volume}
  {80}},\ \bibinfo {pages} {130--150} (\bibinfo {year} {2018})}\BibitemShut
  {NoStop}%
\bibitem [{\citenamefont {Ashwin}, \citenamefont {Perryman},\ and\
  \citenamefont {Wieczorek}(2017)}]{ashwin2017parameter}%
  \BibitemOpen
  \bibfield  {author} {\bibinfo {author} {\bibfnamefont {P.}~\bibnamefont
  {Ashwin}}, \bibinfo {author} {\bibfnamefont {C.}~\bibnamefont {Perryman}}, \
  and\ \bibinfo {author} {\bibfnamefont {S.}~\bibnamefont {Wieczorek}},\
  }\bibfield  {title} {\enquote {\bibinfo {title} {Parameter shifts for
  nonautonomous systems in low dimension: bifurcation-and rate-induced
  tipping},}\ }\href@noop {} {\bibfield  {journal} {\bibinfo  {journal}
  {Nonlinearity}\ }\textbf {\bibinfo {volume} {30}},\ \bibinfo {pages} {2185}
  (\bibinfo {year} {2017})}\BibitemShut {NoStop}%
\bibitem [{\citenamefont {Scharpf}\ \emph {et~al.}(1987)\citenamefont
  {Scharpf}, \citenamefont {Squicciarini}, \citenamefont {Bromley},
  \citenamefont {Green}, \citenamefont {Tredicce},\ and\ \citenamefont
  {Narducci}}]{scharpf1987experimental}%
  \BibitemOpen
  \bibfield  {author} {\bibinfo {author} {\bibfnamefont {W.}~\bibnamefont
  {Scharpf}}, \bibinfo {author} {\bibfnamefont {M.}~\bibnamefont
  {Squicciarini}}, \bibinfo {author} {\bibfnamefont {D.}~\bibnamefont
  {Bromley}}, \bibinfo {author} {\bibfnamefont {C.}~\bibnamefont {Green}},
  \bibinfo {author} {\bibfnamefont {J.}~\bibnamefont {Tredicce}}, \ and\
  \bibinfo {author} {\bibfnamefont {L.}~\bibnamefont {Narducci}},\ }\bibfield
  {title} {\enquote {\bibinfo {title} {Experimental observation of a delayed
  bifurcation at the threshold of an argon laser},}\ }\href@noop {} {\bibfield
  {journal} {\bibinfo  {journal} {Optics communications}\ }\textbf {\bibinfo
  {volume} {63}},\ \bibinfo {pages} {344--348} (\bibinfo {year}
  {1987})}\BibitemShut {NoStop}%
\bibitem [{\citenamefont {Pisarchik}\ \emph {et~al.}(2014)\citenamefont
  {Pisarchik}, \citenamefont {Jaimes-Re{\'a}tegui}, \citenamefont
  {Magall{\'o}n-Garc{\'\i}a},\ and\ \citenamefont
  {Castillo-Morales}}]{pisarchik2014critical}%
  \BibitemOpen
  \bibfield  {author} {\bibinfo {author} {\bibfnamefont {A.~N.}\ \bibnamefont
  {Pisarchik}}, \bibinfo {author} {\bibfnamefont {R.}~\bibnamefont
  {Jaimes-Re{\'a}tegui}}, \bibinfo {author} {\bibfnamefont {C.~A.}\
  \bibnamefont {Magall{\'o}n-Garc{\'\i}a}}, \ and\ \bibinfo {author}
  {\bibfnamefont {C.~O.}\ \bibnamefont {Castillo-Morales}},\ }\bibfield
  {title} {\enquote {\bibinfo {title} {Critical slowing down and noise-induced
  intermittency in bistable perception: bifurcation analysis},}\ }\href@noop {}
  {\bibfield  {journal} {\bibinfo  {journal} {Biological Cybernetics}\ }\textbf
  {\bibinfo {volume} {108}},\ \bibinfo {pages} {397--404} (\bibinfo {year}
  {2014})}\BibitemShut {NoStop}%
\bibitem [{\citenamefont {Matveev}(2003)}]{matveev2003thermoacoustic}%
  \BibitemOpen
  \bibfield  {author} {\bibinfo {author} {\bibfnamefont {K.~I.}\ \bibnamefont
  {Matveev}},\ }\emph {\bibinfo {title} {Thermoacoustic instabilities in the
  Rijke tube: experiments and modeling}},\ \href@noop {} {Ph.D. thesis},\
  \bibinfo  {school} {California Institute of Technology} (\bibinfo {year}
  {2003})\BibitemShut {NoStop}%
\bibitem [{\citenamefont {Juniper}(2011)}]{juniper2011triggering}%
  \BibitemOpen
  \bibfield  {author} {\bibinfo {author} {\bibfnamefont {M.~P.}\ \bibnamefont
  {Juniper}},\ }\bibfield  {title} {\enquote {\bibinfo {title} {Triggering in
  the horizontal rijke tube: non-normality, transient growth and bypass
  transition},}\ }\href@noop {} {\bibfield  {journal} {\bibinfo  {journal}
  {Journal of Fluid Mechanics}\ }\textbf {\bibinfo {volume} {667}},\ \bibinfo
  {pages} {272--308} (\bibinfo {year} {2011})}\BibitemShut {NoStop}%
\bibitem [{\citenamefont {Gopalakrishnan}\ and\ \citenamefont
  {Sujith}(2015)}]{gopalakrishnan2015effect}%
  \BibitemOpen
  \bibfield  {author} {\bibinfo {author} {\bibfnamefont {E.~A.}\ \bibnamefont
  {Gopalakrishnan}}\ and\ \bibinfo {author} {\bibfnamefont {R.~I.}\
  \bibnamefont {Sujith}},\ }\bibfield  {title} {\enquote {\bibinfo {title}
  {Effect of external noise on the hysteresis characteristics of a
  thermoacoustic system},}\ }\href@noop {} {\bibfield  {journal} {\bibinfo
  {journal} {Journal of Fluid Mechanics}\ }\textbf {\bibinfo {volume} {776}},\
  \bibinfo {pages} {334--353} (\bibinfo {year} {2015})}\BibitemShut {NoStop}%
\bibitem [{\citenamefont {Juniper}\ and\ \citenamefont
  {Sujith}(2018)}]{juniper2018sensitivity}%
  \BibitemOpen
  \bibfield  {author} {\bibinfo {author} {\bibfnamefont {M.~P.}\ \bibnamefont
  {Juniper}}\ and\ \bibinfo {author} {\bibfnamefont {R.~I.}\ \bibnamefont
  {Sujith}},\ }\bibfield  {title} {\enquote {\bibinfo {title} {Sensitivity and
  nonlinearity of thermoacoustic oscillations},}\ }\href@noop {} {\bibfield
  {journal} {\bibinfo  {journal} {Annual Review of Fluid Mechanics}\ }\textbf
  {\bibinfo {volume} {50}},\ \bibinfo {pages} {661--689} (\bibinfo {year}
  {2018})}\BibitemShut {NoStop}%
\bibitem [{\citenamefont {Fisher}\ and\ \citenamefont
  {Rahman}(2009)}]{fisher2009remembering}%
  \BibitemOpen
  \bibfield  {author} {\bibinfo {author} {\bibfnamefont {S.~C.}\ \bibnamefont
  {Fisher}}\ and\ \bibinfo {author} {\bibfnamefont {S.~A.}\ \bibnamefont
  {Rahman}},\ }\bibfield  {title} {\enquote {\bibinfo {title} {Remembering the
  giants: Apollo rocket propulsion development},}\ }\href@noop {} {\  (\bibinfo
  {year} {2009})}\BibitemShut {NoStop}%
\bibitem [{\citenamefont {Lieuwen}\ and\ \citenamefont
  {Yang}(2005)}]{lieuwen2005combustion}%
  \BibitemOpen
  \bibfield  {author} {\bibinfo {author} {\bibfnamefont {T.~C.}\ \bibnamefont
  {Lieuwen}}\ and\ \bibinfo {author} {\bibfnamefont {V.}~\bibnamefont {Yang}},\
  }\href@noop {} {\emph {\bibinfo {title} {Combustion instabilities in gas
  turbine engines: operational experience, fundamental mechanisms, and
  modeling}}}\ (\bibinfo  {publisher} {American Institute of Aeronautics and
  Astronautics},\ \bibinfo {year} {2005})\BibitemShut {NoStop}%
\bibitem [{\citenamefont {Fleming}(1998)}]{fleming1998turbine}%
  \BibitemOpen
  \bibfield  {author} {\bibinfo {author} {\bibfnamefont {C.}~\bibnamefont
  {Fleming}},\ }\bibfield  {title} {\enquote {\bibinfo {title} {Turbine makers
  are caught in innovation trap},}\ }\href@noop {} {\bibfield  {journal}
  {\bibinfo  {journal} {Wall Street Journal}\ } (\bibinfo {year} {February 13,
  1998})}\BibitemShut {NoStop}%
\bibitem [{\citenamefont {Strogatz}\ \emph {et~al.}(1994)\citenamefont
  {Strogatz}, \citenamefont {Friedman}, \citenamefont {Mallinckrodt},
  \citenamefont {McKay} \emph {et~al.}}]{strogatz1994nonlinear}%
  \BibitemOpen
  \bibfield  {author} {\bibinfo {author} {\bibfnamefont {S.~H.}\ \bibnamefont
  {Strogatz}}, \bibinfo {author} {\bibfnamefont {M.}~\bibnamefont {Friedman}},
  \bibinfo {author} {\bibfnamefont {A.~J.}\ \bibnamefont {Mallinckrodt}},
  \bibinfo {author} {\bibfnamefont {S.}~\bibnamefont {McKay}},  \emph
  {et~al.},\ }\bibfield  {title} {\enquote {\bibinfo {title} {Nonlinear
  dynamics and chaos: With applications to physics, biology, chemistry, and
  engineering},}\ }\href@noop {} {\bibfield  {journal} {\bibinfo  {journal}
  {Computers in Physics}\ }\textbf {\bibinfo {volume} {8}},\ \bibinfo {pages}
  {532--532} (\bibinfo {year} {1994})}\BibitemShut {NoStop}%
\bibitem [{\citenamefont {Dakos}\ \emph {et~al.}(2012)\citenamefont {Dakos},
  \citenamefont {Van~Nes}, \citenamefont {D'Odorico},\ and\ \citenamefont
  {Scheffer}}]{dakos2012robustness}%
  \BibitemOpen
  \bibfield  {author} {\bibinfo {author} {\bibfnamefont {V.}~\bibnamefont
  {Dakos}}, \bibinfo {author} {\bibfnamefont {E.~H.}\ \bibnamefont {Van~Nes}},
  \bibinfo {author} {\bibfnamefont {P.}~\bibnamefont {D'Odorico}}, \ and\
  \bibinfo {author} {\bibfnamefont {M.}~\bibnamefont {Scheffer}},\ }\bibfield
  {title} {\enquote {\bibinfo {title} {Robustness of variance and
  autocorrelation as indicators of critical slowing down},}\ }\href@noop {}
  {\bibfield  {journal} {\bibinfo  {journal} {Ecology}\ }\textbf {\bibinfo
  {volume} {93}},\ \bibinfo {pages} {264--271} (\bibinfo {year}
  {2012})}\BibitemShut {NoStop}%
\bibitem [{\citenamefont {Lenton}\ \emph {et~al.}(2009)\citenamefont {Lenton},
  \citenamefont {Myerscough}, \citenamefont {Marsh}, \citenamefont {Livina},
  \citenamefont {Price}, \citenamefont {Cox},\ and\ \citenamefont
  {GENIE-team}}]{lenton2009using}%
  \BibitemOpen
  \bibfield  {author} {\bibinfo {author} {\bibfnamefont {T.~M.}\ \bibnamefont
  {Lenton}}, \bibinfo {author} {\bibfnamefont {R.~J.}\ \bibnamefont
  {Myerscough}}, \bibinfo {author} {\bibfnamefont {R.}~\bibnamefont {Marsh}},
  \bibinfo {author} {\bibfnamefont {V.~N.}\ \bibnamefont {Livina}}, \bibinfo
  {author} {\bibfnamefont {A.~R.}\ \bibnamefont {Price}}, \bibinfo {author}
  {\bibfnamefont {S.~J.}\ \bibnamefont {Cox}}, \ and\ \bibinfo {author}
  {\bibnamefont {GENIE-team}},\ }\bibfield  {title} {\enquote {\bibinfo {title}
  {Using {GENIE} to study a tipping point in the climate system},}\ }\href@noop
  {} {\bibfield  {journal} {\bibinfo  {journal} {Philosophical Transactions of
  the Royal Society A: Mathematical, Physical and Engineering Sciences}\
  }\textbf {\bibinfo {volume} {367}},\ \bibinfo {pages} {871--884} (\bibinfo
  {year} {2009})}\BibitemShut {NoStop}%
\bibitem [{\citenamefont {Guttal}\ and\ \citenamefont
  {Jayaprakash}(2008)}]{guttal2008changing}%
  \BibitemOpen
  \bibfield  {author} {\bibinfo {author} {\bibfnamefont {V.}~\bibnamefont
  {Guttal}}\ and\ \bibinfo {author} {\bibfnamefont {C.}~\bibnamefont
  {Jayaprakash}},\ }\bibfield  {title} {\enquote {\bibinfo {title} {Changing
  skewness: an early warning signal of regime shifts in ecosystems},}\
  }\href@noop {} {\bibfield  {journal} {\bibinfo  {journal} {Ecology letters}\
  }\textbf {\bibinfo {volume} {11}},\ \bibinfo {pages} {450--460} (\bibinfo
  {year} {2008})}\BibitemShut {NoStop}%
\bibitem [{\citenamefont {Nair}\ and\ \citenamefont
  {Sujith}(2014)}]{nair2014multifractality}%
  \BibitemOpen
  \bibfield  {author} {\bibinfo {author} {\bibfnamefont {V.}~\bibnamefont
  {Nair}}\ and\ \bibinfo {author} {\bibfnamefont {R.~I.}\ \bibnamefont
  {Sujith}},\ }\bibfield  {title} {\enquote {\bibinfo {title} {Multifractality
  in combustion noise: predicting an impending combustion instability},}\
  }\href@noop {} {\bibfield  {journal} {\bibinfo  {journal} {Journal of Fluid
  Mechanics}\ }\textbf {\bibinfo {volume} {747}},\ \bibinfo {pages} {635--655}
  (\bibinfo {year} {2014})}\BibitemShut {NoStop}%
\bibitem [{\citenamefont {Sujith}\ and\ \citenamefont
  {Unni}(2020)}]{sujith2020complex}%
  \BibitemOpen
  \bibfield  {author} {\bibinfo {author} {\bibfnamefont {R.~I.}\ \bibnamefont
  {Sujith}}\ and\ \bibinfo {author} {\bibfnamefont {V.~R.}\ \bibnamefont
  {Unni}},\ }\bibfield  {title} {\enquote {\bibinfo {title} {Complex system
  approach to investigate and mitigate thermoacoustic instability in turbulent
  combustors},}\ }\href@noop {} {\bibfield  {journal} {\bibinfo  {journal}
  {Physics of Fluids}\ }\textbf {\bibinfo {volume} {32}},\ \bibinfo {pages}
  {061401} (\bibinfo {year} {2020})}\BibitemShut {NoStop}%
\bibitem [{\citenamefont {Mandelbrot}(1983)}]{mandelbrot1983fractal}%
  \BibitemOpen
  \bibfield  {author} {\bibinfo {author} {\bibfnamefont {B.~B.}\ \bibnamefont
  {Mandelbrot}},\ }\href@noop {} {\emph {\bibinfo {title} {The fractal geometry
  of nature}}},\ Vol.\ \bibinfo {volume} {173}\ (\bibinfo  {publisher} {WH
  freeman New York},\ \bibinfo {year} {1983})\BibitemShut {NoStop}%
\bibitem [{\citenamefont {Hurst}(1951)}]{hurst1951long}%
  \BibitemOpen
  \bibfield  {author} {\bibinfo {author} {\bibfnamefont {H.~E.}\ \bibnamefont
  {Hurst}},\ }\bibfield  {title} {\enquote {\bibinfo {title} {Long-term storage
  capacity of reservoirs},}\ }\href@noop {} {\bibfield  {journal} {\bibinfo
  {journal} {Trans. Amer. Soc. Civil Eng.}\ }\textbf {\bibinfo {volume}
  {116}},\ \bibinfo {pages} {770--799} (\bibinfo {year} {1951})}\BibitemShut
  {NoStop}%
\bibitem [{\citenamefont {Kantelhardt}\ \emph {et~al.}(2002)\citenamefont
  {Kantelhardt}, \citenamefont {Zschiegner}, \citenamefont {Koscielny-Bunde},
  \citenamefont {Havlin}, \citenamefont {Bunde},\ and\ \citenamefont
  {Stanley}}]{kantelhardt2002multifractal}%
  \BibitemOpen
  \bibfield  {author} {\bibinfo {author} {\bibfnamefont {J.~W.}\ \bibnamefont
  {Kantelhardt}}, \bibinfo {author} {\bibfnamefont {S.~A.}\ \bibnamefont
  {Zschiegner}}, \bibinfo {author} {\bibfnamefont {E.}~\bibnamefont
  {Koscielny-Bunde}}, \bibinfo {author} {\bibfnamefont {S.}~\bibnamefont
  {Havlin}}, \bibinfo {author} {\bibfnamefont {A.}~\bibnamefont {Bunde}}, \
  and\ \bibinfo {author} {\bibfnamefont {H.~E.}\ \bibnamefont {Stanley}},\
  }\bibfield  {title} {\enquote {\bibinfo {title} {Multifractal detrended
  fluctuation analysis of nonstationary time series},}\ }\href@noop {}
  {\bibfield  {journal} {\bibinfo  {journal} {Physica A: Statistical Mechanics
  and its Applications}\ }\textbf {\bibinfo {volume} {316}},\ \bibinfo {pages}
  {87--114} (\bibinfo {year} {2002})}\BibitemShut {NoStop}%
\bibitem [{\citenamefont {Ivanov}\ \emph {et~al.}(1999)\citenamefont {Ivanov},
  \citenamefont {Amaral}, \citenamefont {Goldberger}, \citenamefont {Havlin},
  \citenamefont {Rosenblum}, \citenamefont {Struzik},\ and\ \citenamefont
  {Stanley}}]{ivanov1999multifractality}%
  \BibitemOpen
  \bibfield  {author} {\bibinfo {author} {\bibfnamefont {P.~C.}\ \bibnamefont
  {Ivanov}}, \bibinfo {author} {\bibfnamefont {L.~A.~N.}\ \bibnamefont
  {Amaral}}, \bibinfo {author} {\bibfnamefont {A.~L.}\ \bibnamefont
  {Goldberger}}, \bibinfo {author} {\bibfnamefont {S.}~\bibnamefont {Havlin}},
  \bibinfo {author} {\bibfnamefont {M.~G.}\ \bibnamefont {Rosenblum}}, \bibinfo
  {author} {\bibfnamefont {Z.~R.}\ \bibnamefont {Struzik}}, \ and\ \bibinfo
  {author} {\bibfnamefont {H.~E.}\ \bibnamefont {Stanley}},\ }\bibfield
  {title} {\enquote {\bibinfo {title} {Multifractality in human heartbeat
  dynamics},}\ }\href@noop {} {\bibfield  {journal} {\bibinfo  {journal}
  {Nature}\ }\textbf {\bibinfo {volume} {399}},\ \bibinfo {pages} {461}
  (\bibinfo {year} {1999})}\BibitemShut {NoStop}%
\bibitem [{\citenamefont {Hu}\ \emph {et~al.}(2004)\citenamefont {Hu},
  \citenamefont {Ivanov}, \citenamefont {Hilton}, \citenamefont {Chen},
  \citenamefont {Ayers}, \citenamefont {Stanley},\ and\ \citenamefont
  {Shea}}]{hu2004endogenous}%
  \BibitemOpen
  \bibfield  {author} {\bibinfo {author} {\bibfnamefont {K.}~\bibnamefont
  {Hu}}, \bibinfo {author} {\bibfnamefont {P.~C.}\ \bibnamefont {Ivanov}},
  \bibinfo {author} {\bibfnamefont {M.~F.}\ \bibnamefont {Hilton}}, \bibinfo
  {author} {\bibfnamefont {Z.}~\bibnamefont {Chen}}, \bibinfo {author}
  {\bibfnamefont {R.~T.}\ \bibnamefont {Ayers}}, \bibinfo {author}
  {\bibfnamefont {H.~E.}\ \bibnamefont {Stanley}}, \ and\ \bibinfo {author}
  {\bibfnamefont {S.~A.}\ \bibnamefont {Shea}},\ }\bibfield  {title} {\enquote
  {\bibinfo {title} {Endogenous circadian rhythm in an index of cardiac
  vulnerability independent of changes in behavior},}\ }\href@noop {}
  {\bibfield  {journal} {\bibinfo  {journal} {Proceedings of the National
  Academy of Sciences}\ }\textbf {\bibinfo {volume} {101}},\ \bibinfo {pages}
  {18223--18227} (\bibinfo {year} {2004})}\BibitemShut {NoStop}%
\bibitem [{\citenamefont {Grech}\ and\ \citenamefont
  {Mazur}(2004)}]{grech2004can}%
  \BibitemOpen
  \bibfield  {author} {\bibinfo {author} {\bibfnamefont {D.}~\bibnamefont
  {Grech}}\ and\ \bibinfo {author} {\bibfnamefont {Z.}~\bibnamefont {Mazur}},\
  }\bibfield  {title} {\enquote {\bibinfo {title} {Can one make any crash
  prediction in finance using the local {H}urst exponent idea?}}\ }\href@noop
  {} {\bibfield  {journal} {\bibinfo  {journal} {Physica A: Statistical
  Mechanics and its Applications}\ }\textbf {\bibinfo {volume} {336}},\
  \bibinfo {pages} {133--145} (\bibinfo {year} {2004})}\BibitemShut {NoStop}%
\bibitem [{\citenamefont {Vandewalle}\ and\ \citenamefont
  {Ausloos}(1997)}]{vandewalle1997coherent}%
  \BibitemOpen
  \bibfield  {author} {\bibinfo {author} {\bibfnamefont {N.}~\bibnamefont
  {Vandewalle}}\ and\ \bibinfo {author} {\bibfnamefont {M.}~\bibnamefont
  {Ausloos}},\ }\bibfield  {title} {\enquote {\bibinfo {title} {Coherent and
  random sequences in financial fluctuations},}\ }\href@noop {} {\bibfield
  {journal} {\bibinfo  {journal} {Physica A: Statistical Mechanics and its
  Applications}\ }\textbf {\bibinfo {volume} {246}},\ \bibinfo {pages}
  {454--459} (\bibinfo {year} {1997})}\BibitemShut {NoStop}%
\bibitem [{\citenamefont {Grech}\ and\ \citenamefont
  {Pamu{\l}a}(2008)}]{grech2008local}%
  \BibitemOpen
  \bibfield  {author} {\bibinfo {author} {\bibfnamefont {D.}~\bibnamefont
  {Grech}}\ and\ \bibinfo {author} {\bibfnamefont {G.}~\bibnamefont
  {Pamu{\l}a}},\ }\bibfield  {title} {\enquote {\bibinfo {title} {The local
  {H}urst exponent of the financial time series in the vicinity of crashes on
  the polish stock exchange market},}\ }\href@noop {} {\bibfield  {journal}
  {\bibinfo  {journal} {Physica A: statistical mechanics and its applications}\
  }\textbf {\bibinfo {volume} {387}},\ \bibinfo {pages} {4299--4308} (\bibinfo
  {year} {2008})}\BibitemShut {NoStop}%
\bibitem [{\citenamefont {Alvarez-Ramirez}\ \emph {et~al.}(2008)\citenamefont
  {Alvarez-Ramirez}, \citenamefont {Alvarez}, \citenamefont {Rodriguez},\ and\
  \citenamefont {Fernandez-Anaya}}]{alvarez2008time}%
  \BibitemOpen
  \bibfield  {author} {\bibinfo {author} {\bibfnamefont {J.}~\bibnamefont
  {Alvarez-Ramirez}}, \bibinfo {author} {\bibfnamefont {J.}~\bibnamefont
  {Alvarez}}, \bibinfo {author} {\bibfnamefont {E.}~\bibnamefont {Rodriguez}},
  \ and\ \bibinfo {author} {\bibfnamefont {G.}~\bibnamefont
  {Fernandez-Anaya}},\ }\bibfield  {title} {\enquote {\bibinfo {title}
  {Time-varying {H}urst exponent for us stock markets},}\ }\href@noop {}
  {\bibfield  {journal} {\bibinfo  {journal} {Physica A: Statistical Mechanics
  and its Applications}\ }\textbf {\bibinfo {volume} {387}},\ \bibinfo {pages}
  {6159--6169} (\bibinfo {year} {2008})}\BibitemShut {NoStop}%
\bibitem [{\citenamefont {Matos}\ \emph {et~al.}(2008)\citenamefont {Matos},
  \citenamefont {Gama}, \citenamefont {Ruskin}, \citenamefont {Al~Sharkasi},\
  and\ \citenamefont {Crane}}]{matos2008time}%
  \BibitemOpen
  \bibfield  {author} {\bibinfo {author} {\bibfnamefont {J.~A.}\ \bibnamefont
  {Matos}}, \bibinfo {author} {\bibfnamefont {S.~M.}\ \bibnamefont {Gama}},
  \bibinfo {author} {\bibfnamefont {H.~J.}\ \bibnamefont {Ruskin}}, \bibinfo
  {author} {\bibfnamefont {A.}~\bibnamefont {Al~Sharkasi}}, \ and\ \bibinfo
  {author} {\bibfnamefont {M.}~\bibnamefont {Crane}},\ }\bibfield  {title}
  {\enquote {\bibinfo {title} {Time and scale {H}urst exponent analysis for
  financial markets},}\ }\href@noop {} {\bibfield  {journal} {\bibinfo
  {journal} {Physica A: Statistical Mechanics and its Applications}\ }\textbf
  {\bibinfo {volume} {387}},\ \bibinfo {pages} {3910--3915} (\bibinfo {year}
  {2008})}\BibitemShut {NoStop}%
\bibitem [{\citenamefont {Domino}(2011)}]{domino2011use}%
  \BibitemOpen
  \bibfield  {author} {\bibinfo {author} {\bibfnamefont {K.}~\bibnamefont
  {Domino}},\ }\bibfield  {title} {\enquote {\bibinfo {title} {The use of the
  {H}urst exponent to predict changes in trends on the warsaw stock
  exchange},}\ }\href@noop {} {\bibfield  {journal} {\bibinfo  {journal}
  {Physica A: Statistical Mechanics and its Applications}\ }\textbf {\bibinfo
  {volume} {390}},\ \bibinfo {pages} {98--109} (\bibinfo {year}
  {2011})}\BibitemShut {NoStop}%
\bibitem [{\citenamefont {Suyal}, \citenamefont {Prasad},\ and\ \citenamefont
  {Singh}(2009)}]{suyal2009nonlinear}%
  \BibitemOpen
  \bibfield  {author} {\bibinfo {author} {\bibfnamefont {V.}~\bibnamefont
  {Suyal}}, \bibinfo {author} {\bibfnamefont {A.}~\bibnamefont {Prasad}}, \
  and\ \bibinfo {author} {\bibfnamefont {H.~P.}\ \bibnamefont {Singh}},\
  }\bibfield  {title} {\enquote {\bibinfo {title} {Nonlinear time series
  analysis of sunspot data},}\ }\href@noop {} {\bibfield  {journal} {\bibinfo
  {journal} {Solar Physics}\ }\textbf {\bibinfo {volume} {260}},\ \bibinfo
  {pages} {441--449} (\bibinfo {year} {2009})}\BibitemShut {NoStop}%
\bibitem [{\citenamefont {Kilcik}\ \emph {et~al.}(2009)\citenamefont {Kilcik},
  \citenamefont {Anderson}, \citenamefont {Rozelot}, \citenamefont {Ye},
  \citenamefont {Sugihara},\ and\ \citenamefont {Ozguc}}]{kilcik2009nonlinear}%
  \BibitemOpen
  \bibfield  {author} {\bibinfo {author} {\bibfnamefont {A.}~\bibnamefont
  {Kilcik}}, \bibinfo {author} {\bibfnamefont {C.~N.~K.}\ \bibnamefont
  {Anderson}}, \bibinfo {author} {\bibfnamefont {J.~P.}\ \bibnamefont
  {Rozelot}}, \bibinfo {author} {\bibfnamefont {H.}~\bibnamefont {Ye}},
  \bibinfo {author} {\bibfnamefont {G.}~\bibnamefont {Sugihara}}, \ and\
  \bibinfo {author} {\bibfnamefont {A.}~\bibnamefont {Ozguc}},\ }\bibfield
  {title} {\enquote {\bibinfo {title} {Nonlinear prediction of solar cycle
  24},}\ }\href@noop {} {\bibfield  {journal} {\bibinfo  {journal} {The
  Astrophysical Journal}\ }\textbf {\bibinfo {volume} {693}},\ \bibinfo {pages}
  {1173} (\bibinfo {year} {2009})}\BibitemShut {NoStop}%
\bibitem [{\citenamefont {Unni}\ and\ \citenamefont
  {Sujith}(2015)}]{unni2015multifractal}%
  \BibitemOpen
  \bibfield  {author} {\bibinfo {author} {\bibfnamefont {V.~R.}\ \bibnamefont
  {Unni}}\ and\ \bibinfo {author} {\bibfnamefont {R.~I.}\ \bibnamefont
  {Sujith}},\ }\bibfield  {title} {\enquote {\bibinfo {title} {Multifractal
  characteristics of combustor dynamics close to lean blowout},}\ }\href@noop
  {} {\bibfield  {journal} {\bibinfo  {journal} {Journal of Fluid Mechanics}\
  }\textbf {\bibinfo {volume} {784}},\ \bibinfo {pages} {30--50} (\bibinfo
  {year} {2015})}\BibitemShut {NoStop}%
\bibitem [{\citenamefont {Gotoda}\ \emph {et~al.}(2012)\citenamefont {Gotoda},
  \citenamefont {Amano}, \citenamefont {Miyano}, \citenamefont {Ikawa},
  \citenamefont {Maki},\ and\ \citenamefont
  {Tachibana}}]{gotoda2012characterization}%
  \BibitemOpen
  \bibfield  {author} {\bibinfo {author} {\bibfnamefont {H.}~\bibnamefont
  {Gotoda}}, \bibinfo {author} {\bibfnamefont {M.}~\bibnamefont {Amano}},
  \bibinfo {author} {\bibfnamefont {T.}~\bibnamefont {Miyano}}, \bibinfo
  {author} {\bibfnamefont {T.}~\bibnamefont {Ikawa}}, \bibinfo {author}
  {\bibfnamefont {K.}~\bibnamefont {Maki}}, \ and\ \bibinfo {author}
  {\bibfnamefont {S.}~\bibnamefont {Tachibana}},\ }\bibfield  {title} {\enquote
  {\bibinfo {title} {Characterization of complexities in combustion instability
  in a lean premixed gas-turbine model combustor},}\ }\href@noop {} {\bibfield
  {journal} {\bibinfo  {journal} {Chaos: An Interdisciplinary Journal of
  Nonlinear Science}\ }\textbf {\bibinfo {volume} {22}},\ \bibinfo {pages}
  {043128} (\bibinfo {year} {2012})}\BibitemShut {NoStop}%
\bibitem [{\citenamefont {Havlin}\ \emph {et~al.}(1999)\citenamefont {Havlin},
  \citenamefont {Amaral}, \citenamefont {Ashkenazy}, \citenamefont
  {Goldberger}, \citenamefont {Ivanov}, \citenamefont {Peng},\ and\
  \citenamefont {Stanley}}]{havlin1999application}%
  \BibitemOpen
  \bibfield  {author} {\bibinfo {author} {\bibfnamefont {S.}~\bibnamefont
  {Havlin}}, \bibinfo {author} {\bibfnamefont {L.~A.~N.}\ \bibnamefont
  {Amaral}}, \bibinfo {author} {\bibfnamefont {Y.}~\bibnamefont {Ashkenazy}},
  \bibinfo {author} {\bibfnamefont {A.~L.}\ \bibnamefont {Goldberger}},
  \bibinfo {author} {\bibfnamefont {P.~C.}\ \bibnamefont {Ivanov}}, \bibinfo
  {author} {\bibfnamefont {C.-K.}\ \bibnamefont {Peng}}, \ and\ \bibinfo
  {author} {\bibfnamefont {H.~E.}\ \bibnamefont {Stanley}},\ }\bibfield
  {title} {\enquote {\bibinfo {title} {Application of statistical physics to
  heartbeat diagnosis},}\ }\href@noop {} {\bibfield  {journal} {\bibinfo
  {journal} {Physica A: Statistical Mechanics and its Applications}\ }\textbf
  {\bibinfo {volume} {274}},\ \bibinfo {pages} {99--110} (\bibinfo {year}
  {1999})}\BibitemShut {NoStop}%
\bibitem [{\citenamefont {Telesca}\ \emph {et~al.}(2001)\citenamefont
  {Telesca}, \citenamefont {Cuomo}, \citenamefont {Lapenna},\ and\
  \citenamefont {Macchiato}}]{telesca2001new}%
  \BibitemOpen
  \bibfield  {author} {\bibinfo {author} {\bibfnamefont {L.}~\bibnamefont
  {Telesca}}, \bibinfo {author} {\bibfnamefont {V.}~\bibnamefont {Cuomo}},
  \bibinfo {author} {\bibfnamefont {V.}~\bibnamefont {Lapenna}}, \ and\
  \bibinfo {author} {\bibfnamefont {M.}~\bibnamefont {Macchiato}},\ }\bibfield
  {title} {\enquote {\bibinfo {title} {A new approach to investigate the
  correlation between geoelectrical time fluctuations and earthquakes in a
  seismic area of southern italy},}\ }\href@noop {} {\bibfield  {journal}
  {\bibinfo  {journal} {Geophysical Research Letters}\ }\textbf {\bibinfo
  {volume} {28}},\ \bibinfo {pages} {4375--4378} (\bibinfo {year}
  {2001})}\BibitemShut {NoStop}%
\bibitem [{\citenamefont {Qian}\ and\ \citenamefont
  {Rasheed}(2004)}]{qian2004hurst}%
  \BibitemOpen
  \bibfield  {author} {\bibinfo {author} {\bibfnamefont {B.}~\bibnamefont
  {Qian}}\ and\ \bibinfo {author} {\bibfnamefont {K.}~\bibnamefont {Rasheed}},\
  }\bibfield  {title} {\enquote {\bibinfo {title} {Hurst exponent and financial
  market predictability},}\ }in\ \href@noop {} {\emph {\bibinfo {booktitle}
  {IASTED conference on Financial Engineering and Applications}}}\ (\bibinfo
  {year} {2004})\ pp.\ \bibinfo {pages} {203--209}\BibitemShut {NoStop}%
\bibitem [{\citenamefont {Gopalakrishnan}\ \emph {et~al.}(2016)\citenamefont
  {Gopalakrishnan}, \citenamefont {Sharma}, \citenamefont {John}, \citenamefont
  {Dutta},\ and\ \citenamefont {Sujith}}]{gopalakrishnan2016early}%
  \BibitemOpen
  \bibfield  {author} {\bibinfo {author} {\bibfnamefont {E.~A.}\ \bibnamefont
  {Gopalakrishnan}}, \bibinfo {author} {\bibfnamefont {Y.}~\bibnamefont
  {Sharma}}, \bibinfo {author} {\bibfnamefont {T.}~\bibnamefont {John}},
  \bibinfo {author} {\bibfnamefont {P.~S.}\ \bibnamefont {Dutta}}, \ and\
  \bibinfo {author} {\bibfnamefont {R.~I.}\ \bibnamefont {Sujith}},\ }\bibfield
   {title} {\enquote {\bibinfo {title} {Early warning signals for critical
  transitions in a thermoacoustic system},}\ }\href@noop {} {\bibfield
  {journal} {\bibinfo  {journal} {Scientific reports}\ }\textbf {\bibinfo
  {volume} {6}},\ \bibinfo {pages} {1--10} (\bibinfo {year}
  {2016})}\BibitemShut {NoStop}%
\bibitem [{\citenamefont {{Ghanavati}}\ \emph {et~al.}(2014)\citenamefont
  {{Ghanavati}}, \citenamefont {{Hines}}, \citenamefont {{Lakoba}},\ and\
  \citenamefont {{Cotilla-Sanchez}}}]{6848858}%
  \BibitemOpen
  \bibfield  {author} {\bibinfo {author} {\bibfnamefont {G.}~\bibnamefont
  {{Ghanavati}}}, \bibinfo {author} {\bibfnamefont {P.~D.~H.}\ \bibnamefont
  {{Hines}}}, \bibinfo {author} {\bibfnamefont {T.~I.}\ \bibnamefont
  {{Lakoba}}}, \ and\ \bibinfo {author} {\bibfnamefont {E.}~\bibnamefont
  {{Cotilla-Sanchez}}},\ }\bibfield  {title} {\enquote {\bibinfo {title}
  {Understanding early indicators of critical transitions in power systems from
  autocorrelation functions},}\ }\href@noop {} {\bibfield  {journal} {\bibinfo
  {journal} {IEEE Transactions on Circuits and Systems I: Regular Papers}\
  }\textbf {\bibinfo {volume} {61}},\ \bibinfo {pages} {2747--2760} (\bibinfo
  {year} {2014})}\BibitemShut {NoStop}%
\bibitem [{\citenamefont {Wiesenfeld}(1985)}]{wiesenfeld1985noisy}%
  \BibitemOpen
  \bibfield  {author} {\bibinfo {author} {\bibfnamefont {K.}~\bibnamefont
  {Wiesenfeld}},\ }\bibfield  {title} {\enquote {\bibinfo {title} {Noisy
  precursors of nonlinear instabilities},}\ }\href@noop {} {\bibfield
  {journal} {\bibinfo  {journal} {Journal of Statistical Physics}\ }\textbf
  {\bibinfo {volume} {38}},\ \bibinfo {pages} {1071--1097} (\bibinfo {year}
  {1985})}\BibitemShut {NoStop}%
\bibitem [{\citenamefont {Noiray}(2017)}]{noiray2017linear}%
  \BibitemOpen
  \bibfield  {author} {\bibinfo {author} {\bibfnamefont {N.}~\bibnamefont
  {Noiray}},\ }\bibfield  {title} {\enquote {\bibinfo {title} {Linear growth
  rate estimation from dynamics and statistics of acoustic signal envelope in
  turbulent combustors},}\ }\href@noop {} {\bibfield  {journal} {\bibinfo
  {journal} {Journal of Engineering for Gas Turbines and Power}\ }\textbf
  {\bibinfo {volume} {139}},\ \bibinfo {pages} {041503} (\bibinfo {year}
  {2017})}\BibitemShut {NoStop}%
\bibitem [{\citenamefont {Fujisaka}\ and\ \citenamefont
  {Inoue}(1989)}]{fujisaka1989correlation}%
  \BibitemOpen
  \bibfield  {author} {\bibinfo {author} {\bibfnamefont {H.}~\bibnamefont
  {Fujisaka}}\ and\ \bibinfo {author} {\bibfnamefont {M.}~\bibnamefont
  {Inoue}},\ }\bibfield  {title} {\enquote {\bibinfo {title} {Correlation
  functions of temporal fluctuations},}\ }\href@noop {} {\bibfield  {journal}
  {\bibinfo  {journal} {Physical Review A}\ }\textbf {\bibinfo {volume} {39}},\
  \bibinfo {pages} {4778} (\bibinfo {year} {1989})}\BibitemShut {NoStop}%
\bibitem [{\citenamefont {Varfolomeyev}\ and\ \citenamefont
  {Gurevich}(2001)}]{varfolomeyev2001hyperexponential}%
  \BibitemOpen
  \bibfield  {author} {\bibinfo {author} {\bibfnamefont {S.~D.}\ \bibnamefont
  {Varfolomeyev}}\ and\ \bibinfo {author} {\bibfnamefont {K.~G.}\ \bibnamefont
  {Gurevich}},\ }\bibfield  {title} {\enquote {\bibinfo {title} {The
  hyperexponential growth of the human population on a macrohistorical
  scale},}\ }\href@noop {} {\bibfield  {journal} {\bibinfo  {journal} {Journal
  of Theoretical Biology}\ }\textbf {\bibinfo {volume} {212}},\ \bibinfo
  {pages} {367--372} (\bibinfo {year} {2001})}\BibitemShut {NoStop}%
\bibitem [{\citenamefont {Ihlen}(2012)}]{ihlen2012introduction}%
  \BibitemOpen
  \bibfield  {author} {\bibinfo {author} {\bibfnamefont {E.~A. F.~E.}\
  \bibnamefont {Ihlen}},\ }\bibfield  {title} {\enquote {\bibinfo {title}
  {Introduction to multifractal detrended fluctuation analysis in {M}atlab},}\
  }\href@noop {} {\bibfield  {journal} {\bibinfo  {journal} {Frontiers in
  physiology}\ }\textbf {\bibinfo {volume} {3}},\ \bibinfo {pages} {141}
  (\bibinfo {year} {2012})}\BibitemShut {NoStop}%
\end{thebibliography}%

\end{document}